\documentclass[reprint,amsmath,amssymb,aps]{revtex4-2}
\usepackage[utf8]{inputenc}
\usepackage{graphicx}
\usepackage{amsmath}
\usepackage{xspace}
\newcommand{\beacon}{BEACON\xspace}

\begin{document}

\title{Global optimization of atomic structures with gradient-enhanced Gaussian process regression}
\author{Sami Kaappa}
\affiliation{Department of Physics, Technical University of Denmark, Kongens Lyngby, Denmark}

\author{Estefan\'ia Garijo del R\'io}
\affiliation{Department of Physics, Technical University of Denmark, Kongens Lyngby, Denmark}

\author{Karsten Wedel Jacobsen}
\affiliation{Department of Physics, Technical University of Denmark, Kongens Lyngby, Denmark}
\email{kwj@fysik.dtu.dk}
\date{\today}

\newlength{\figwidth}
\setlength{\figwidth}{0.95\columnwidth}
\newlength{\widefig}
\setlength{\widefig}{0.8\textwidth}

\begin{abstract}
Determination of atomic structures is a key challenge in the fields of computational physics and materials science, as a large variety of mechanical, chemical, electronic, and optical properties depend sensitively on structure. Here, we present a global optimization scheme where energy and force information from density functional theory (DFT) calculations is transferred to a probabilistic surrogate model to estimate both the potential energy surface (PES) and the associated uncertainties. The local minima in the surrogate PES are then used to guide the search for the global minimum in the DFT potential. We find that adding the gradients in most cases improves the efficiency of the search significantly. The method is applied to global optimization of [Ta$_2$O$_5$]$_x$ clusters with $x=1,2,3$, and the surface structure of oxidized ZrN.
\end{abstract}

\maketitle

\section{Introduction}
Global optimization in high-dimensional space is a long-standing challenge in numerical analysis, and also in physics, chemistry, and material science. The structure of an atomic system is at low temperature given by the global minimum point of the potential energy surface (PES), which is a function, $E(\mathbf{x})$, of all atomic coordinates $\mathbf{x}$. For atomic systems with more than a few atoms, the dimensionality therefore constitutes a challenge. Furthermore, the PES is usually determined by a quantum mechanical calculation with for example density functional theory (DFT), and these calculations are computationally demanding so the optimization should therefore be performed with as few function evaluations as possible.

Numerous algorithms for finding the structural ground state of a system are implemented for material science problems \cite{Zhang2020}, such as basin hopping \cite{basinhopping}, evolutionary algorithms, \cite{EA2, sioclusters, Jaeger2019}, particle swarm optimization \cite{SGO}, and random searches \cite{randomsearch}, but the issue with these methods remains the large number of DFT evaluations required by the algorithms.

Recently, machine-learned surrogate models have been considered in order to overcome the problem of spending excessive amounts of computer resources on DFT calculations. A surrogate model for the PES is constructed based on a dataset typically obtained with DFT, and it allows for subsequent much faster evaluation of atomic energies and forces. Surrogate models have been used in local optimization \cite{GPMin}, global optimization \cite{pmlr-v48-carr16, BO-rrss-2018, Mortensen2020reinforced, jager2018machine, sioclusters, Malthe2020efficient, Rinderspacher2020, doi:10.1021/acs.jctc.0c00648}, nudged-elastic band calculations \cite{koinstinen2017, AID2020Garrido}, searches for transition states \cite{denzel2018transition}, adsorption studies \cite{todorovic2019bayesian}, and the design of force fields \cite{GAP, Chmiela2016FF, Vandermause2020}. 

Many of the surrogate models are based on Bayesian inference, or Gaussian processes (GP) \cite{williams2006gaussian, Bishop2006}, where the resulting PES is a sum of joint kernel functions, centered at the training points. In its most traditional form, the GP is only trained with the target values, i.e. the electronic ground-state energies in the context of computational chemistry. However, since forces are readily available after a ground state DFT calculation, we train the model also with the gradients of the target value, i.e. with the forces on each atom. The inclusion of gradients is crucial in local structure optimization based on surrogate models \cite{GPMin} and has also been shown to generally improve model predictions \cite{christensen2020gradients}. 

The construction of the PES based on a GP usually involves the introduction of a distance measure or similarity of two different atomic configurations. If two configurations are close, it is assumed that energies and forces will be close as well. To this end it may be advantageous to describe the atomic configurations using a structural fingerprint (alias descriptor), which in the simplest case that we shall consider here, is simply a mapping from the atomic Cartesian coordinates to a (typically high-dimensional) vector. The similarity of two atomic configurations can then be estimated based on the difference between the two fingerprint vectors.

The introduction of a fingerprint vector may have several advantages.  For example, a fingerprint can be constructed to reflect the translational, rotational, and permutational invariances of the atomic configuration, \emph{i.e.} if two configurations differ by only a permutation of identical atoms, the fingerprint will be unchanged. This has the consequence that the predicted PES will exhibit the same symmetries.  Furthermore, a good descriptor is able to catch the relevant information of the configuration for the underlying problem. A simple example of an atomic fingerprint is the Coulomb matrix \cite{lilienfeld2012coulomb}, which represents the atomic configuration using inverse distances, but more elaborate fingerprints have been developed during the latest years, such as SOAP \cite{SOAP}, ACSF \cite{ACSF}, many-body tensor representation \cite{MBTR}, and FCHL \cite{FCHL}. Most of the common fingerprints are ready-to-use in DScribe package \cite{dscribe}, although currently the gradients of the descriptions are not available, that are highly relevant in optimization problems.

In this work, we use Bayesian optimization \cite{7352306} in order to find global minimum structures for various systems. The work follows the pioneering approach for efficient global optimization of atomic structures by Bisbo and Hammer \cite{Malthe2020efficient,Bisbo:2020vo} with the essential difference that we train our GP regression model with both energies and forces. We also note that the implementation of the approach involves many choices and parameters, where we might differ from Bisbo and Hammer. For example, we use a similar global fingerprint, but introduce an additional smooth cutoff function to obtain a smoother representation of the gradients. In general, the gradients are seen to improve the efficiency of the global optimization, and we illustrate this through applications to a 15-atom Cu cluster, bulk SiO$_2$, Ti$_4$O$_8$ cluster, bulk TiO$_2$ and bulk silicon.

This article is organized as follows. In section \ref{sec:surrogatemodel}, we describe the surrogate model that we use to predict energies and forces of atomic structures during a global search. In section \ref{sec:valid}, we illustrate the predictive power of the model by generating learning curves for a Cu$_{15}$ cluster and bulk SiO$_2$. In section \ref{sec:glopt}, we describe the global optimization approach, and then in section \ref{sec:goresults} we demonstrate its performance on a Cu$_{15}$ cluster, bulk SiO$_2$, [Ta$_2$O$_5$]$_x$ clusters, a ZrN-O surface, a Ti$_4$O$_8$ cluster, bulk TiO$_2$, and bulk silicon. In section \ref{sec:compperf}, we discuss computational performance issues before we finally conclude.

\section{Surrogate model}\label{sec:surrogatemodel}
\subsection{Gaussian process with gradients}\label{subsec:gp}
To model the potential energy surface of an atomic structure, we use a Gaussian process that learns energies and forces (\emph{i.e.} negative gradients) from existing data. A Gaussian process uses Bayesian inference and is based on the assumption that the prior distribution for the data is given by a multi-dimensional normal distribution. The result is that the predicted energy and forces, $\mu(\mathbf{x}) = (E(\mathbf{x}),-\mathbf{F}(\mathbf{x}))$, at a given atomic configuration $\mathbf x$  with fingerprint $\rho(\mathbf x)$ can be described \cite{williams2006gaussian, poloczek2017gradients} as
\begin{equation}\label{gp_mean}
    \mu(\mathbf x) = \mu_p(\mathbf x) + K(\rho(\mathbf x), P)C(P, P)^{-1} (y-\mu_p(X)),
\end{equation}
where $X$ is a list with the atomic configurations of the training data and $P=\rho(X)$ a list with the corresponding fingerprints. $y$ is a vector that contains energies and negative forces of the training data, $\mu_p$ denotes the prior energy and negative forces ($\mu_p(\mathbf x)$ for the given configuration and $\mu_p(X)$ for the training data points), $K$ denotes the covariance matrix, and $C$ is the regularized $K$-matrix of covariances between training data points. The forces are inserted as their negatives because the mathematical expression of a Gaussian process works with the gradients and not with the forces. The resulting vector $\mu(\mathbf x)$ contains the predicted total energy and the negative predicted forces for each atom (in Cartesian coordinates). For the sake of readability, we will denote $\rho=\rho(\mathbf x)$ for the fingerprint for the rest of this section. 

The covariance matrix $K$ is built as follows. A covariance matrix $\tilde K$ between two atomic configurations is written as \cite{poloczek2017gradients}
\begin{equation}\label{Kmatrix}
    \tilde K(\rho_1, \rho_2) = \left( \begin{matrix}
    k(\rho_1, \rho_2) & \left(\nabla_2 k(\rho_1, \rho_2)\right)^T\\
    \nabla_{1} k(\rho_1, \rho_2) &  \nabla_{1}\left( \nabla_2 k(\rho_1, \rho_2)\right)^T
    \end{matrix}\right)
\end{equation}
with kernel (or covariance) function $k(\rho_1, \rho_2)$. Here, $\nabla_i$ operates on the Cartesian coordinates of the atomic configuration $\mathbf x_i$, represented by the fingerprint $\rho_i$. The components in the matrix on the right-hand side of Eq. \ref{Kmatrix} can be interpreted as the covariances between energies ($k$), covariances between energies and forces ($\nabla_i k$) and covariances between different force components ($\nabla_i \nabla_j k$). 

Using the formulation of $\tilde K$, we store the covariances between a single fingerprint $\rho$ and all the training point fingerprints in $P$ in the matrix $K(\rho, P)$ as
\begin{equation}
    K_{j}(\rho, P) = \tilde K(\rho, P_j)
\end{equation}
where $j$ runs over the number of training points. Finally, the matrix $C$ contains the covariance between all the training points and a diagonal regularization that describes the estimated noise, or uncertainty, of the training data. Its elements are thus given by

\begin{equation}
    C_{ij}(P, P) = \tilde{K}(P_i, P_j) + \mathrm{diag}\left(\sigma_{n,E}^2, \sigma_{n,f}^2\right)\delta_{ij}
\end{equation}
where $\text{diag}(\sigma_{n,E}^2, \sigma_{n,f}^2)$ 
represents the diagonal matrix with $\sigma_{n,E}^2$ in the first entry and $\sigma_{n,f}^2$ in the remaining ones. In this way, we have introduced separate regularizations $\sigma_{n,E}^2$ and $\sigma_{n,f}^2$ for energy and force covariances, respectively. Throughout this work, the regularization is set so that the ratios between the regularization parameters and the kernel prefactor $\sigma$ (defined below in equation \ref{eq:sqexp}) are $\sigma_{n,E}/\sigma= 0.0005$ for energies and $\sigma_{n,f}/\sigma=0.001$ for forces.

The power of using Bayesian inference in searching the global minimum comes from the estimated uncertainties of the predictions that are easily attainable. For the Gaussian process, the estimated standard deviation is given by \cite{williams2006gaussian}
\begin{equation}
    \Sigma(\mathbf x) = \left[\tilde K(\rho, \rho) - K(\rho, P)C(P, P)^{-1} K(P, \rho) \right] ^{1/2}
\end{equation}
The uncertainties are used for the global optimization through the acquisition function as described below in section \ref{sec:glopt}.

\subsection{The model}\label{subsec:fp}
We describe the atomic structure by a fingerprint which has two terms: a radial distribution and an angular distribution. Using such global distributions ensure the rotational, translational, and permutational symmetries for a system. The radial distribution is motivated by the radial distribution function described by Valle and Oganov \cite{oganovfp}. However, to remove discontinuity at the cutoff distance, we use a smooth weighting factor. The radial part of the fingerprint for element pair $AB$ is calculated as
\begin{equation}\label{fp_radial}
 { \rho}_{AB}^R(r; {\mathbf x}) = \sum_{\substack{i\in A \\ j\in B}} \frac{1}{r_{ij}^2}f_c(r_{ij}; R_c^R) \, e^{-|r-r_{ij}|^2/2\delta_R^2}
\end{equation}
where $r_{ij}$ is distance between atoms $i$ and $j$ in the set of coordinates $\mathbf x$, and $\delta_R$ is a smearing factor with a fixed value $\delta_R=0.4$ \AA. $A$ and $B$ indicate different elements in the system, and the $i$-sum goes only over atoms that are of element $A$ and the $j$-sum goes only over atoms that are of element $B$. $r$ denotes the discrete variable with 200 values, ranging from 0 to the cutoff distance $R_c^R$. The smooth function $f_c$ has the form
\begin{equation}
    f_c(r; R_c) = 
    \begin{cases}
    1 - (1+\gamma)\left(\frac{r}{R_c}\right)^\gamma + \gamma\left(\frac{r}{R_c}\right)^{1+\gamma} & \mathrm{if~} r \leq R_c\\
    0 & \mathrm{if~} r > R_c
    \end{cases}
\end{equation}
with a cutoff distance $R_c$ and $\gamma=2$. This form for $f_c(r)$ has zero value and zero derivative at $r=R_c$. Due to the factor 1/$r^2$, the form of equation \ref{fp_radial} has the property of giving more weight on small distances below $R_c^R$. The angular part of the fingerprint is given by
\begin{equation}
    { \rho}_{ABC}^\alpha(\theta; \mathbf x) = \sum_{\substack{ i\in A \\ j\in B \\k\in C}}f_c(r_{ij}; R_c^\alpha)f_c(r_{jk}; R_c^\alpha) \, e^{-|\theta-\theta_{ijk}|^2/2\delta_\alpha^2}
\end{equation}
where $\theta$ is a discrete variable with 100 values that range from 0 to $\pi$, $\theta_{ijk}$ is the angle between atoms $i$, $j$ and $k$, and $\delta_\alpha$ is a smearing factor with a fixed value $\delta_\alpha=0.4 \mathrm{~rad}$. The chosen values for $\delta_R$ and $\delta_\alpha$ were determined by trying a few different values. The ones chosen were observed to work well for all the systems studied in this work. In the smooth function $f_c$ we use the value of $\gamma=0.5$ that again ensures a smooth behavior of the fingerprint at cutoff $R_c=R_c^\alpha$. The different $\gamma$ value to that of the radial part comes from our observation that the predicted potential energy surfaces were not smooth enough at the angular cutoff radius when using the value of $\gamma=2$ in the angular part. In our work, the cutoff radius $R_c^R$ has values between 4.0 and 8.0 \AA, and $R_c^\alpha$ has values between 3.6 and 4.0 \AA, comparable to the radii studied in, e.g., \cite{Vandermause2020}.
The fingerprint is very similar to the one used by Bisbo and Hammer \cite{Bisbo:2020vo} but with the additional cut-off function for the radial part.

The total fingerprint for an atomic configuration $\mathbf x$ is obtained by concatenating all vectors $\rho_{AB}^R(r;\mathbf x)$ and $ \rho_{ABC}^\alpha(\theta; \mathbf x)$ with elements $A$, $B$ and $C$ of the system, resulting in a single vector that we denote as $\rho$. To clarify, for a single-element system such as a Cu cluster, the radial part is just $\rho^R=\rho_\mathrm{Cu,Cu}^R(r;\mathbf x)$, and for a two-element system, like SiO$_2$, the radial fingerprint consists of vectors $\rho_\mathrm{Si,Si}^R(r;\mathbf x)$, $\rho_\mathrm{Si,O}^R(r;\mathbf x)$, $\rho_\mathrm{O, Si}^R(r;\mathbf x)$, and $\rho_\mathrm{O, O}^R(r;\mathbf x)$. A similar procedure is used for the angular parts of the fingerprint.

We calculate the covariance between data points using a squared-exponential kernel 
\begin{equation}\label{eq:sqexp}
    k(\rho_1, \rho_2) = \sigma^2\exp\left(\frac{-D({\rho}_1, {\rho}_2)^2}{2l^2}\right)
\end{equation}
with the distance function $D(\rho_1, \rho_2)$ and two descriptive hyperparameters, the prefactor $\sigma$ and length scale $l$. (Note, that we use the term prefactor for $\sigma$ and not $\sigma^2$.) It is good to note that, although the dimensionality of the fingerprint can be thousands, the kernel function only includes distances between the fingerprint vectors. Therefore, the efficiency of a Gaussian process does not suffer from the high dimensionality of the fingerprint. 

The distance function we take as simply the Euclidean distance between the fingerprint vectors as 
\begin{equation}
    D(\rho_1, \rho_2) = \left[\sum_i (\rho_{1i} - \rho_{2i})^2\right]^{1/2}.
\end{equation}
Since the gradients of the kernel function in Eq. \ref{eq:sqexp} are required by a Gaussian process that is trained on forces (in accordance to Eq. \ref{Kmatrix}),
the full formulas to calculate the gradients with this specific distance function in fingerprint space are given in the Supplemental information \cite{suppinfo}. The forces can be predicted also for the model that is trained on energies only, as we also show in the Supplemental information \cite{suppinfo}.

We note that Bisbo and Hammer \cite{Bisbo:2020vo} use a kernel function, which is a sum of two squared-exponential kernels with two different length scales. We tried this, but did not see any systematic improvement by adding an extra length scale.

We determine the hyperparameters $\sigma$ and $l$, by maximizing the logarithmic marginal likelihood, which is written as \cite{williams2006gaussian}
\begin{align}
    \log \mathcal P = &-\frac{1}{2} \log(\det{C(P, P)}) \nonumber \\
     & - \frac{1}{2}(y-\mu_p(X))^T C(P,P)^{-1} (y-\mu_p(X)) \nonumber \\
     &- \frac{N(3N_\mathrm{atoms} + 1)}{2}\log 2\pi
\end{align}
where $N$ is the number of training points and $N_\mathrm{atoms}$ is the number of atoms in a single training data point. The prefactor, $\sigma$, can be determined analytically for fixed values of $\sigma_{n,E}/\sigma$ and $\sigma_{n,f}/\sigma$, so the numerical optimization problem is only one-dimensional. 

The Gaussian process allows for the specification of a prior function, $E_p(\mathbf x)$, for the energy landscape. In equation (\ref{gp_mean}), the prior energy landscape is inserted as $\mu_p(\mathbf x)=\left(E_p(\mathbf x), \nabla E_p(\mathbf x)\right)$.  Here, we apply the prior function suggested by Bisbo and Hammer \cite{Malthe2020efficient}, which is a repulsive potential of the form
\begin{equation}\label{prior}
    E_p(\mathbf x) = E_c + E_r(\mathbf x) = E_c +\sum_{ij} \left(0.7\,\, \frac{R_i + R_j}{r_{ij}(\mathbf x)}\right)^{12},
\end{equation}
where $E_c$ is a constant, $R_i$ and $R_j$ are the covalent radii of atoms with indices $i$ and $j$, and $r_{ij}(\mathbf x)$ is the distance between the atoms in the set of atomic coordinates $\mathbf x$. The prior energy function expresses the expectation that the energy rises steeply if two atoms come very close. As we shall see later, this helps avoiding very high energy structures in the training data.
 
 The constant value $E_c$ is determined by maximizing the marginal likelihood, and is given by the analytic formula
\begin{equation}\label{prior_constant}
    E_c = \frac{\mathbf U^T C(P, P)^{-1}(y - E_r(X))} {\mathbf U^T C(P, P)^{-1} \mathbf U}
\end{equation}
where $E_r(X)$ is a vector that consists of repulsive priors of the training data, that is obtained using the second term in equation (\ref{prior}), and $\mathbf U$ is a vector of length $N(3N_\mathrm{atoms} + 1)$ with elements
\begin{equation}
    \mathbf U_i=
    \begin{cases}
        1, \quad \mathrm{if} \quad i \, \mathrm{mod} (3N_\mathrm{atoms} + 1) = 0 \\
        0, \quad \mathrm{otherwise}
    \end{cases}
\end{equation} 
where indexing of $i$ starts from 0. Therefore, $\mathbf U_i=1$ if $y_i$ is an energy value, and $\mathbf U_i=0$ if $y_i$ is a force.

\section{Model validation}\label{sec:valid}

\subsection{Learning curves and cross validation}
To validate the model, we examine two different systems: A Cu$_{15}$ cluster using an effective-medium theory (EMT) potential as implemented in the ASE package \cite{EMT, ase-paper, ase} and bulk SiO$_2$ using DFT with the PBE functional \cite{PBE}, implemented in GPAW \cite{gpaw}. The unit cell for SiO$_2$ consists of 12 atoms and has the lowest energy in the cristobalite structure, comprising tetrahedral SiO$_4$ units. For both training and validation sets, random structures are generated and relaxed loosely so that the maximum force criterion is 10 eV/\AA. (The random structures are generated in the same way as in the global optimization runs, which is discussed in detail below, in section \ref{sec:glopt}.) Different sizes of training sets are then used to generate learning curves for models with a variety of length scales in the squared-exponential kernel. We limit the training set size to 100 because with larger sizes there is a risk of memory problems when gradients are trained. The validation set size is kept as 100. 

The reason for showing the learning curves for different length scales is that the length scale has, as we shall see, a dominant effect on the predictive power of the model, and we also observe that the optimal length scales do not necessarily follow the traditional expectations for a Gaussian process. The constant term in the prior function (Equation (\ref{prior})) is set to the mean energy of the training set and the kernel prefactor is kept constant since it does not affect the mean of the predictions, as deduced from equation (\ref{gp_mean}). In addition to fixed length scales, the learning curves are computed for models where the length scale, the prior constant and the kernel prefactor are obtained by maximizing the marginal likelihood separately for each training set size. 

\begin{figure}[ht]
\includegraphics[width=\figwidth]{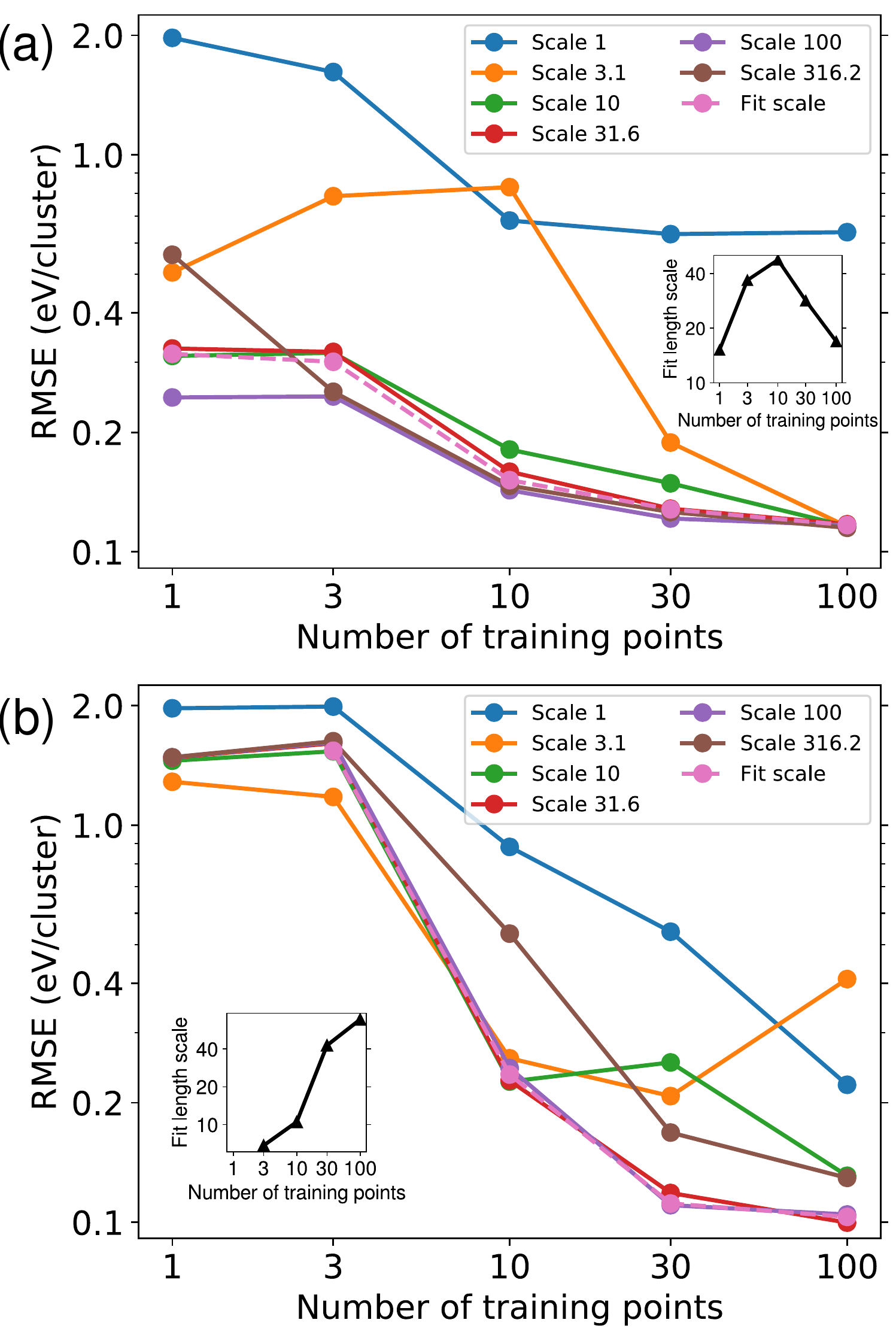}
\caption{\small{Cu$_{15}$ learning curves obtained by (a) training on both energies and forces and (b) training on energies alone. The inset graphs show the values of the fit length scales for each training set size.}}\label{cu15_lc}
\end{figure}

The learning curves for Cu$_{15}$ are shown in Fig. \ref{cu15_lc}  both with and without training the gradients. For the gradient-trained curves, we observe that the root-mean-square error (RMSE) saturates to approximately 0.12 eV/cluster with length scales greater than 30 when the training set size is increased up to 100 data points. The curves with scale 3.1 and 10 do not seem to saturate to the same degree but they end up at the same RMSE at 100 training points as the other models. The standard deviation of the test set energies is 1.5 eV/cluster, meaning that our model can decrease the RMSE to about 7\% of what random sampling of energies would produce. Amazingly, a RMSE of 0.25 eV/cluster is achieved by having only one data point in the training set. This might be due to the ability of the fingerprint to catch the relatively simple radial dependence of the EMT potential, together with the smooth squared-exponential kernel.

Despite the good start with few training points, the curves develop quite slowly with increasing number of training data. The observed saturation is in contrast with the power law behavior that the learning curves should optimally follow with linear learning curves on the log-log scale \cite{Huang2016learning, huang2020ab}. One reason for the saturation could be the difficulty to resolve differences between some configurations as compared to others \cite{Parsaeifard2020}. We have observed that small perturbations of the atomic structure that lead to similar variations in the energy may differ in several orders of magnitude in the variation of the fingerprint. This may indicate that the potential energy surface is highly anisotropic in fingerprint space.  We note that the high dimensionality of the fingerprint also makes correcting for the anisotropy with different scales in each direction impractical. 

Another explanation for the shape of the learning curves could be that the distribution functions used in the fingerprint have a finite range. Deringer and Csányi have shown in the case of amorphous carbon \cite{Deringer2017machine} that a finite cutoff may lead to substantial residual forces for a per-atom fingerprint. We have a global fingerprint, but still the finite range may limit the quality of the model prediction.

In practice, it seems that the gradient-trained model learns everything it can after training of the order of only 30 data points. On the other hand, an accuracy of 0.12 eV/cluster is clearly sufficient to be of relevance for the determination of the basins with low energy configurations.

The effect of training the gradients is apparent in Fig. \ref{cu15_lc}: the prediction error is clearly lower up to training set sizes of the order of 10. At 30 training data points and above, the energy-trained model seems to do as well as the gradient-trained one. With the largest training set sizes, the energy-trained model becomes even slightly better than the one including the gradients. The difference is small, though, with the RMSE difference being only 0.012 eV/cluster at training set size of 100 between the models with the best length scales. However, given the saturation of the gradient-trained curves, it is natural to expect that the energy-trained model reaches the performance of the gradient-model at some point in general, and for this particular system that happens after around 30 training points.  

The models with scales 10 and above saturate to similar performance when the training set size of 100 is reached. All of these scales are comparable to or longer than the distances between the data points in the data set with 100 points. For this data set, the distances vary between 0.1 and 2.3 in fingerprint space. The model with scale 3.1 is seen to perform less well for small data sets, but is not saturated when 100 data points is reached. The limited distances in fingerprint space has to do with the character of the fingerprint. If the atoms in a given configuration are displaced, the Cartesian distance corresponding to the displacement can grow indefinitely. In fingerprint space the distances are calculated based on differences in distribution functions, and these will saturate at some point.

 The length scales obtained by maximizing the log-likelihood are always above 15 when gradients are included in the training, and above 6.8 while training on the energies alone (see Fig. \ref{cu15_lc}). These scales are also surprisingly long considering the distances in the training set. There is a clear increasing trend of the optimal length scale in the energy-trained models when the training set size is increasing, but no trend is observed within the gradient-trained models. Nevertheless, maximizing the log likelihood gives roughly the best model as evaluated with cross validation.

\begin{figure}[ht]
\includegraphics[width=0.96\columnwidth]{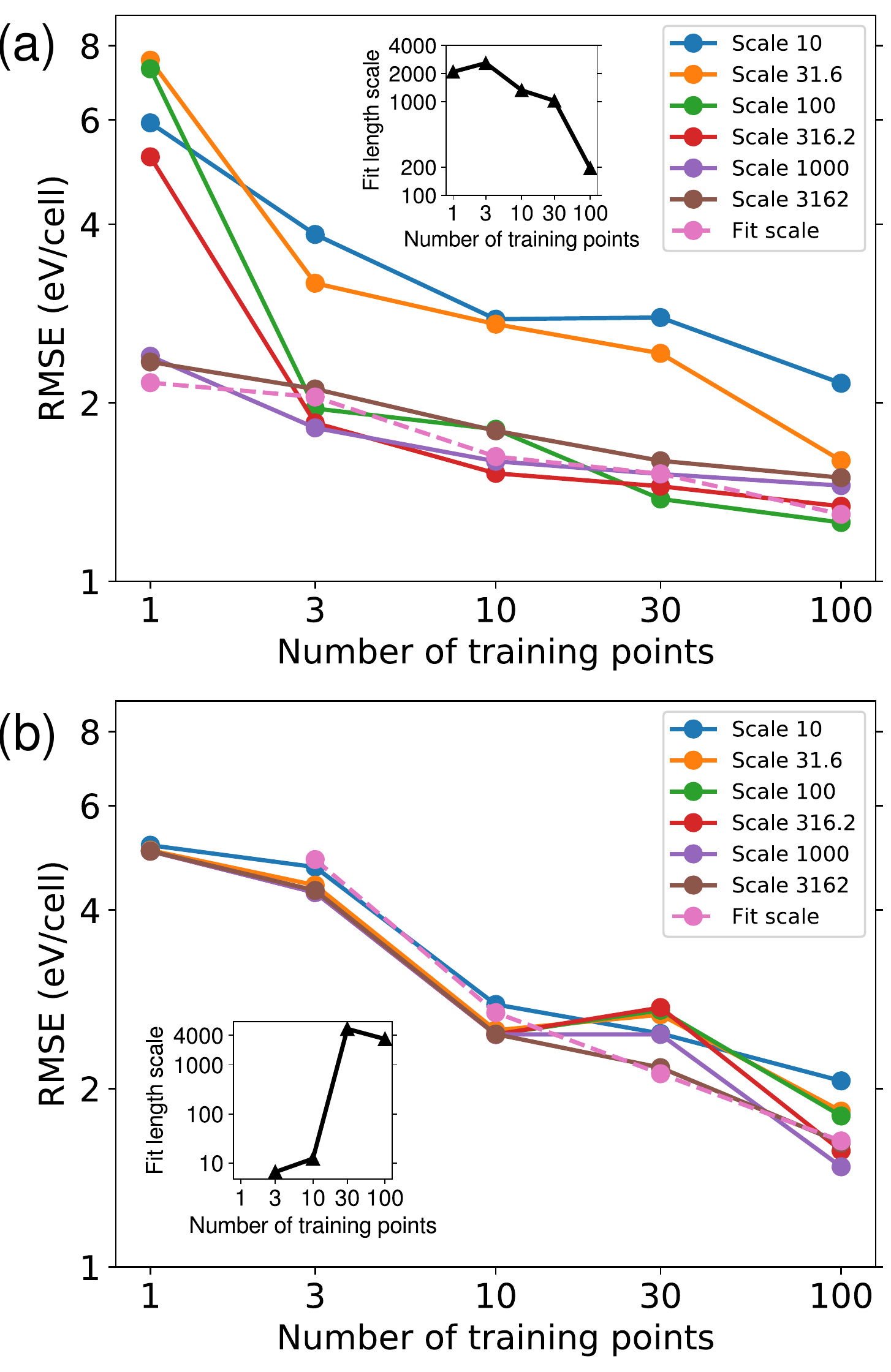}
\caption{\small{SiO$_{2}$ learning curves (a) training energies and forces and (b) training only energies. The inset graphs show the values of the length scales obtained by maximizing the log-likelihood for each training set size.}}\label{sio2_lc}
\end{figure}

For bulk SiO$_2$ (figure \ref{sio2_lc}), the picture is different in that the power law decay of the learning curves is roughly maintained with most models with different length scales up to 100 training points. But, the prediction power is not too impressive with the best RMSE of 1.27 eV/cell at 100 training points although it is lower than the standard deviation of the validation set, which is 3.6 eV/cell. The energy-trained model reaches towards the gradient-trained one again, but at 100 training points the gradient-trained model predicts still slightly better than the model based on energies alone. No clear saturation is observed within this data range with neither of the approaches.

Compared to the EMT cluster, the learning ability with a single training point is reduced. This is expected when moving to a more complex potential energy surface: the quantum effects of a more realistic potential are difficult to track with a relatively simple fingerprint in contrast to the simple radial dependency of EMT.

From the SiO$_2$ learning curves we can also see that the length scales above 100 are clearly favored over the smaller ones. Again, this is remarkable for a Gaussian process with a single squared-exponential kernel as covariance function, since the distances between the test data points vary between 1.6 and 21.3, that is one or two orders of magnitude smaller than the most optimal length scales. According to Fig. \ref{sio2_lc}, the most optimal length scale is as much as 1000 while training only on energies, even further from the deviation of the data point distances compared to the gradient-trained model.

The fitted length scales, obtained by maximizing the log-likelihood and shown in the insets in Fig. \ref{sio2_lc}, have a clear decreasing trend for the models trained using gradients. For the models trained on only energies the variation is dominated by a large jump between 10 and 30 training points. This behaviour indicates a high sensitivity of the model to the addition of more data, since we are dealing with a small numbers of training data points. Another reason might be that there are multiple local maxima in the marginal likelihood, and different maximization runs end up at different local maxima.

Finally, it should be noted that for both Cu$_{15}$ and SiO$_2$, fitting the length scale by maximizing the marginal likelihood gives roughly the best model in the cross validation. This is a desired behavior since maximizing the marginal likelihood is an easy and computationally relatively cheap procedure to carry out, and it can therefore be done repeatedly during a global search where the training set is updated often.

\subsection{Local relaxations in the surrogate model}\label{subsec:localrelax}
Since our global optimization method relies on local relaxations rather than single-point calculations in the surrogate PES, it is relevant to examine how well the model performs local relaxations. We create a training data set of 40 data points of the Cu$_{15}$ cluster, relaxed with EMT so that the maximum force residual is less than 1.0 eV/\AA. After this, the model is trained on the data both with and without gradients. Then, 80 random structures are created independently and relaxed locally in the surrogate model. We note that the minimizations on the surrogate surface are always performed using the predicted forces. This can be done even if the model is trained on energies only, as we note in section \ref{subsec:fp}.

\begin{figure}
    \centering
    \includegraphics[width=\linewidth]{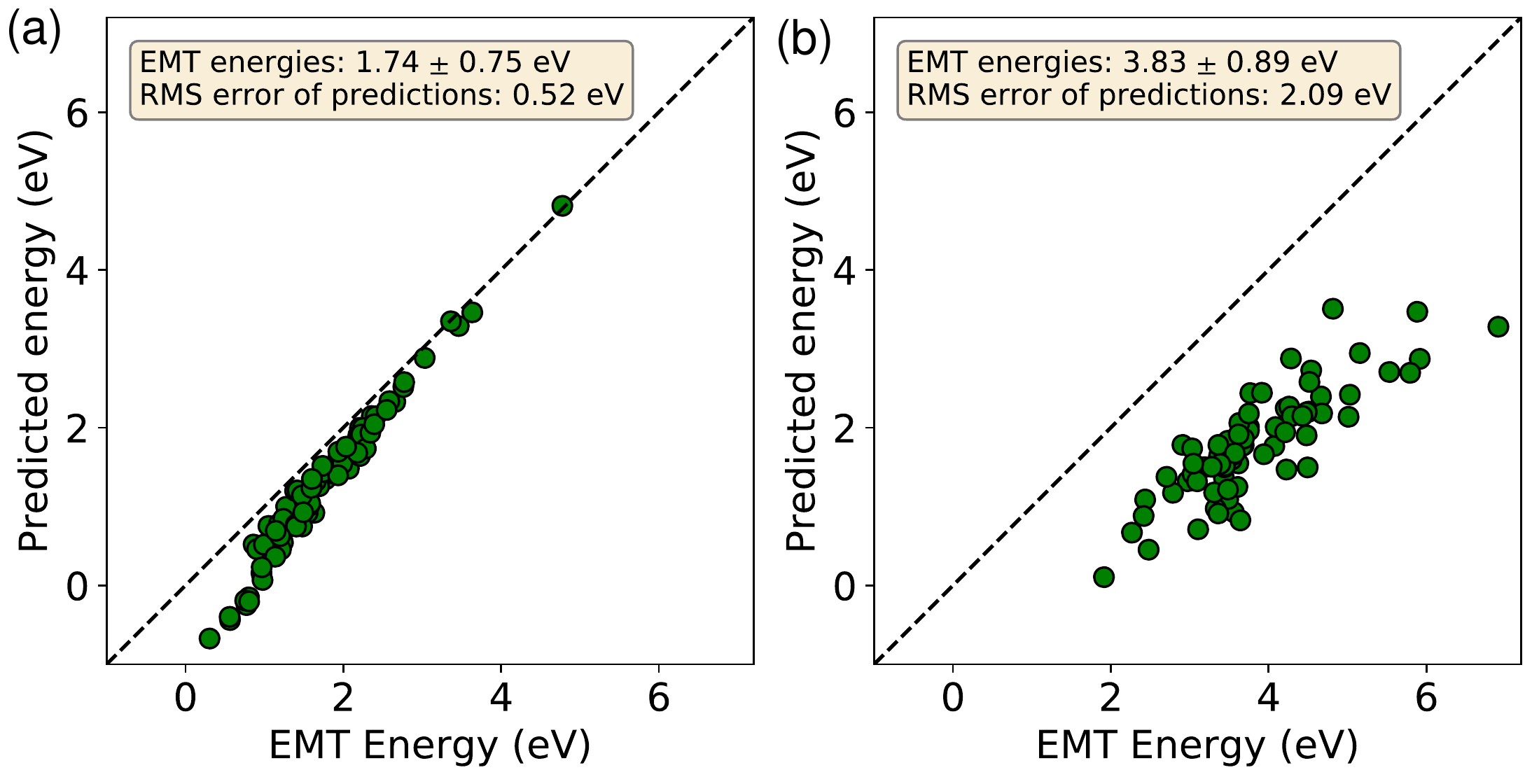}
    \caption{Predicted versus true energies of the final structures of relaxations with surrogate models. 80 local relaxations of Cu$_{15}$ are run with surrogate models with (a) training on the gradients and (b) training only on energies. The training set of 40 data points is the same for both models. The total energies of the training data points have distribution with mean of 6.0 eV and standard deviation of 1.14 eV. The reference point 0.0 eV is set to the global minimum energy. Statistics of the energies of the final structures, as well as the statistical prediction errors, are shown in the text boxes. The dashed line shows the ideal 1-to-1 mapping of the predictions.}
    \label{cu15_fixedpes}
\end{figure}

A comparison between the EMT energies and the energies of the models is shown in Fig. \ref{cu15_fixedpes}. The EMT total energies of the obtained structures range from 0.38 to 4.8 eV with the gradient-trained model, and from 2.0 to 6.9 eV with training only on energies. Actually, the lowest energy structure (of 0.38 eV) corresponds to a structure that is very close to the true global minimum structure. The prediction errors range from -1.0 to 0.0 eV/cluster with gradients and -3.6 to -1.2 eV/cluster without gradients. The data demonstrates that the gradient model is able to reach both lower energies and higher accuracy than the model including only energies, although we work at a training set size of 40 where the learning curves show similar performance to each other. The model trained on gradients exhibits some systematic errors in particular for small energies, which is most probably due to these structures being relatively far from the training data, making the prediction of their energies more difficult. However, the ordering of the energies seems to be well reproduced. The errors are much larger for the model trained on energies alone, but again the ordering of the states is reproduced fairly well.

\section{Global optimization algorithm}\label{sec:glopt}
\begin{figure}
    \centering
    \includegraphics[width=\linewidth]{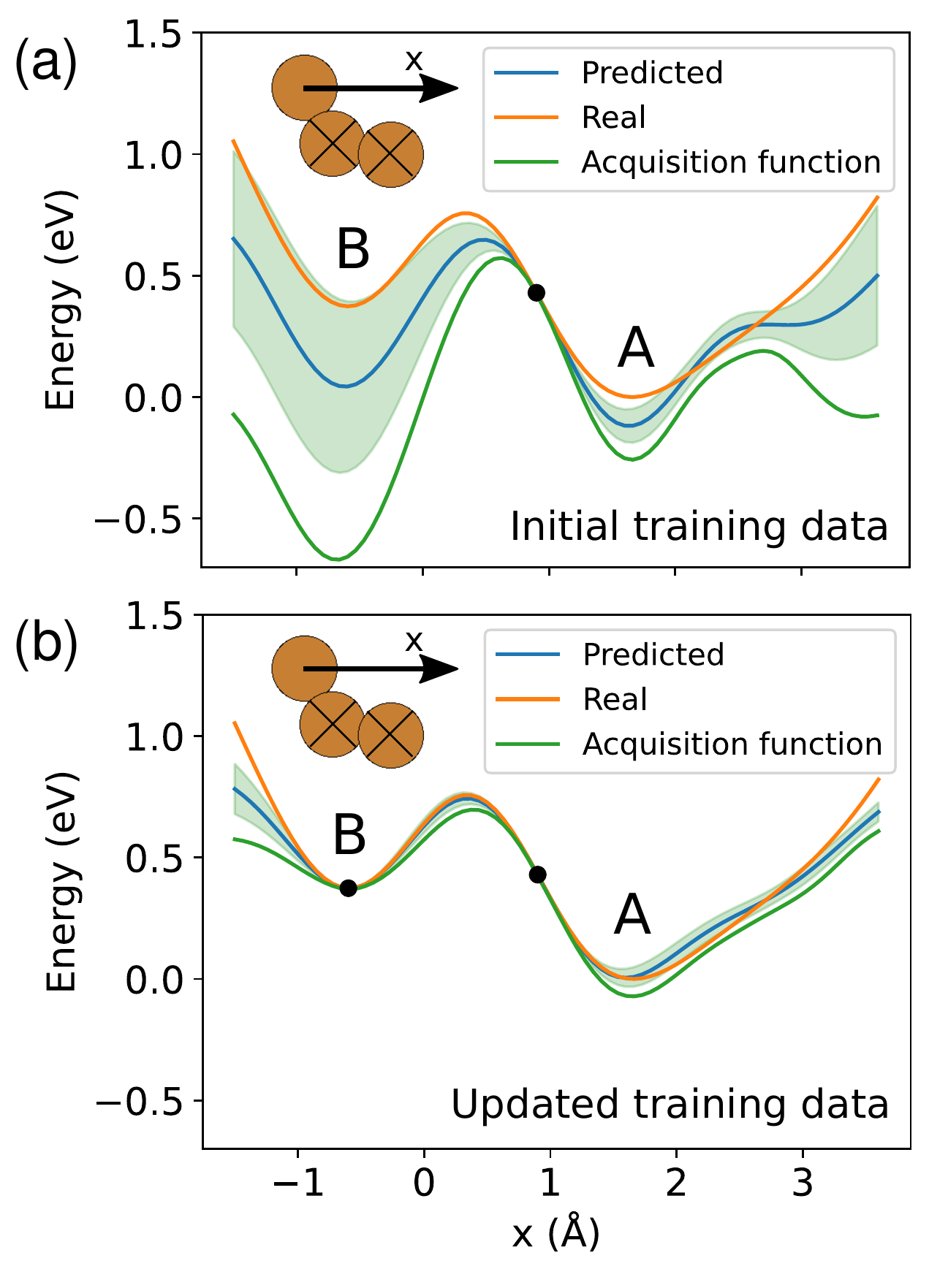}
    \caption{One-dimensional demonstration of surrogate surface. In the figures, the true potential surface, evaluated with EMT, is shown for a system where one Cu atom is moved on top of two other Cu atoms. The two basins of the potential energy surface are labelled as A and B, where the bottom of A has lower energy than B. In (a), a data point is evaluated at $x=0.96$, and a Gaussian process is trained with that single training point. The resulting predicted surface, uncertainty (green area) and acquisition function are shown. The acquisition function is minimized in basin B, at $x=-0.74$, and in (b), we show the predictions after evaluating and adding this point to the training set. Now, the acquisition function is minimized in A, the global minimum basin of the true potential energy surface.}
    \label{demo}
\end{figure}

The algorithm for the global optimization is relatively simple:  in each iterative step, multiple local relaxations are carried out on the surrogate surface, and the energy and forces for the most promising structure are evaluated using the true potential (EMT or DFT). To select the most promising configuration of all the relaxed structures, we make use of the estimated uncertainties by calculating the acquisition function
\begin{equation}\label{acqf}
    f(\mathbf x) = \mu(\mathbf x) - \kappa\Sigma(\mathbf x)
\end{equation}
for each structure, where we set the parameter $\kappa=2$. This form of the acquisition function for minimization problems is called lower confidence bound in the literature, and the choice of $\kappa=2$ provides a good balance between exploitation (low energy) and exploration (large uncertainty) \cite{Jorgensen:2018iu, Malthe2020efficient, Bisbo:2020vo}. The structure with the lowest acquisition function is selected for evaluation with the true potential. This structure is then added to the training set for the Gaussian process and another set of surrogate relaxations are performed with the updated model, as illustrated in Fig. \ref{demo}. The effect of using the fingerprint is visualized in Fig. \ref{demo} as well: after just a single training point, the predicted PES exhibits several important features like the existence of the two local minima. This would not be the case if using Cartesian coordinates as the descriptor due to the missing permutational, rotational and translational symmetries. The gradients also play an important role for quick learning of the features of the PES. Moreover, adding the second training point in one of the basins makes the prediction in the second basin more accurate.

The initial training set consists of randomly generated structures with energies and forces evaluated. In this work, the optimization routine is always started with 2 initial training points. The starting structures for surrogate relaxations comprise three different types: 1) Already visited structures with the lowest energies, 2) already visited structures with random displacement (also called rattling), and 3) randomly generated structures. The total number of surrogate relaxations per step in this work varies between 20 and 40, depending on the system but kept constant during a run. About 25\% of the relaxations start from structures of type 1, 25\% of type 2 and 50\% of type 3. Before accepting the structure given by a surrogate relaxation, it is checked that none of the bond lengths in the system are less than 0.7 times the covalent distance of the atoms. If no valid structures are acquired in an iterative step, a random structure is generated, evaluated and added to the training set, to achieve a model that will fail with smaller probability in the next iteration.

For clusters, the random structures are generated as follows. First, one of the atoms is placed at the origin. After this, the position of the next atom is always given by $\mathbf r_1 = \mathbf r_\mathrm{rand} + (r, \theta, \phi)$ in spherical coordinates where $\mathbf r _\mathrm{rand}$ is the position of one of the previously set atoms, randomly selected, and $(r, \theta, \phi)$ are randomly generated spherical coordinates. Here the sample distribution of $r$ is selected manually and system-specifically, but it was observed that selecting the upper bound of $r$ slightly smaller than the standard covalent distance of the two atoms works best in general. After generating a new position, we make sure that adding an atom in the acquired coordinates does not violate our restriction of the short bond lengths with each atom added to the cluster already, as described above. For TaO clusters, we set another restriction to add some chemical intuition: when O is being added to the system, we enforce that $\mathbf r _\mathrm{rand}$ is the coordinate of a Ta atom and not another O atom. This way we avoid introducing chains and clusters of oxygen in the randomly generated structures.

For bulk systems, the atoms are simply put at random coordinates inside the unit cell, and then relaxed in a repulsive potential of the form of 
\begin{equation}\label{reppot}
    V_\mathrm{rep}(\mathbf x) = \sum_{ij}\left(0.4\,\, \frac{R_i + R_j}{r_{ij}(\mathbf x)}\right)^{12}
\end{equation}
with similar notation to that of the prior in equation \ref{prior}. For surfaces, the atoms are put inside a manually-defined box inside the true unit cell, where the atoms are placed in a similar fashion to that of the bulk systems and then relaxed in the repulsive potential, given by eq. \eqref{reppot}.

The hyperparameters $E_c$, $\sigma$, $\ell$ are updated during the global optimization assuming fixed values of the noise parameters relative to the prefactor as explained above. The prior constant $E_c$ and the prefactor $\sigma$ are obtained analytically and are updated at every step of the global optimization. The update of the length scale has to be done numerically and is performed every five steps with an initial value of 20 times the distance in fingerprint space between the two initial structures. It is our experience that the length scale obtained from maximizing the log-likelihood may be too short leading to a too rapidly varying surrogate PES with more local minima than the true PES. This affects the search so that it becomes too local. This is unfortunate, especially in the beginning of the search, where large parts of the configuration space has to be explored. We therefore introduce a lower bound on the value of the length scale during update. The lower bound is set to the mean value of all distances between the training data points in fingerprint space. Using this value ensures that a large number of training data are used in each energy/force prediction and too local searches are avoided. We also note that the investigations of the learning curves above indicate that a length scale considerably longer than the one obtained from maximizing the log-likelihood still results in a reasonable model. In some cases the prediction error is in fact reduced by increasing the length scale. 

The algorithm is implemented so that the user can choose whether to train using the gradients or not. In both cases, the energy and force predictions can be obtained analytically from the surrogate potential energy surface. The approach where the gradients are not included in the training has a reduced memory usage and the time to train the model is also significantly reduced. However, as we have seen in the investigations of the learning curves the models without training on gradients are less accurate. In the GOFEE method \cite{Malthe2020efficient}, training is only performed on the energies, but another training point, adjacent to the one selected by the acquisition function, is always evaluated with the true potential and added to the training set. The second training point is obtained by moving the atoms a small distance along the direction of the forces. Adding the neighboring data point allows the GP model to have the information about the amplitude of the gradient in the PES in one direction, presumably leading to more accurate predictions. 

In this paper, we refer to our approach with training on both energies an forces as \beacon (from Bayesian Exploration of Atomic Configurations for OptimizatioN) and the approach where the training is only on energies as L-\beacon, where L stands for ``light''. Although the forces are not trained in L-BEACON, we can still predict the forces in the system (as noted in section \ref{subsec:fp}) to be used in the relaxations. We also show results of L-\beacon-exact where a neighboring data point is evaluated and added to the training set similarly to GOFEE, and L-\beacon-FD, where, for each DFT-evaluated data point, we add a neighboring data point where the energy is obtained by a finite difference estimation based on the DFT forces. The step length in the finite difference method is discussed in section \ref{sec:goresults}. We will see that for the systems we investigate here,  adding extra displaced training points does not lead to significant improvement even if gradients are not trained. Furthermore, we find that training only the energies of single points (L-\beacon) makes a surrogate potential energy surface that has similar or almost similar performance in global optimization, compared to L-\beacon-FD and L-\beacon-exact.

As usual in global optimization, Bayesian optimization gives no information whether the true global minimum is achieved, unless the full search space of interest is explored. In BEACON, the search is continued until a given number of DFT calculations is performed, or if time or memory resources are run out. Therefore, the best indication of whether the true global minimum is found is that several separate BEACON runs end up with the same lowest-energy structure.

We also note that the algorithm of BEACON does not include any geometry relaxations on the true PES, but all the relaxations are done on the surrogate PES. In this work, DFT relaxations are only performed if explicitly stated.

\section{Results and discussion}\label{sec:goresults}

\subsection{Cu$_{15}$}
\begin{figure}[ht]
\includegraphics[width=\figwidth]{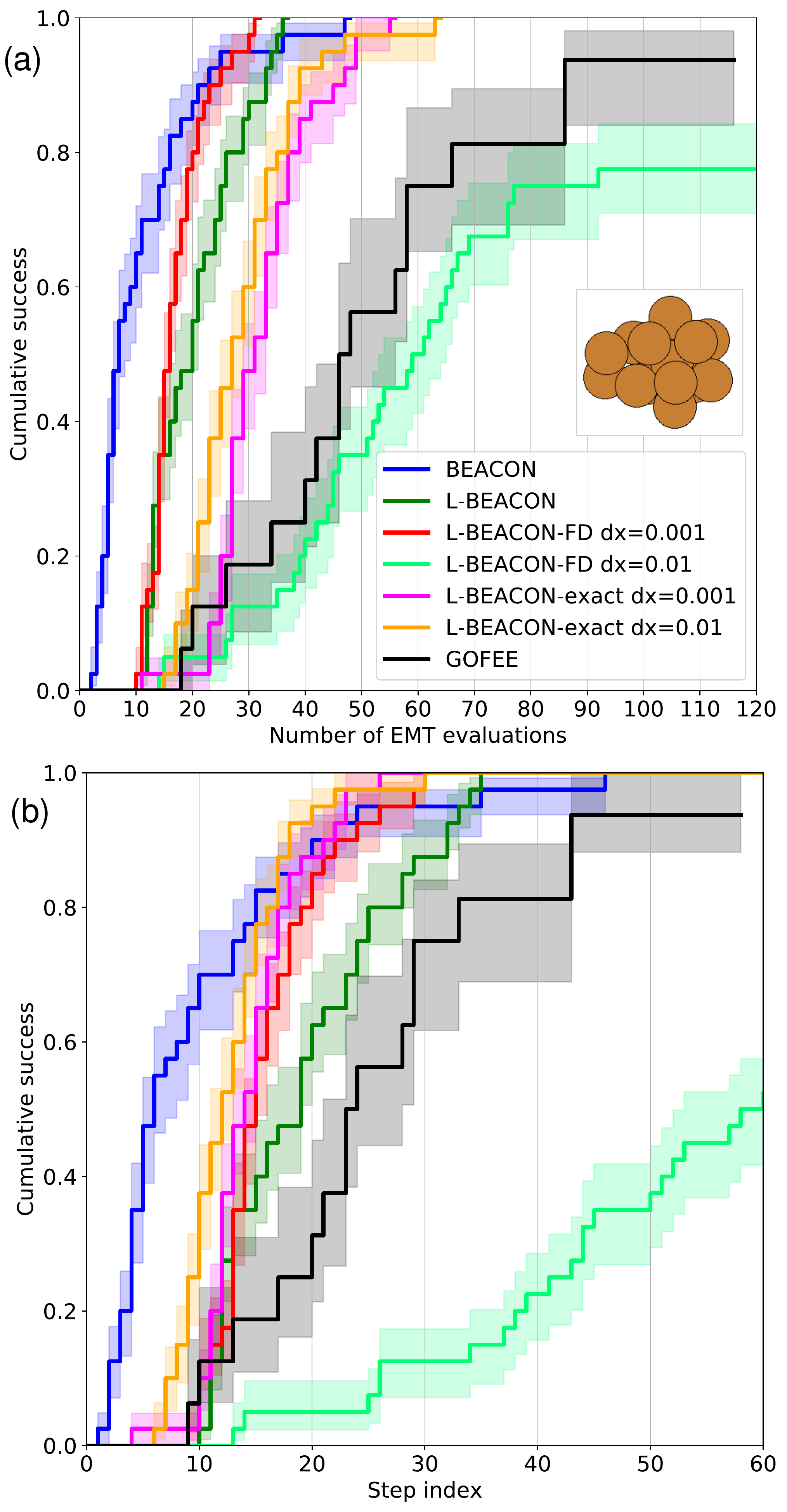}
\caption{\small{Cu$_{15}$ success curves with success threshold of 0.01 eV/cluster. The colored areas denote the standard deviation of bootstrap simulations with 1000 samples of each curve. (a) Success curves as function of number of EMT calculations. (b) Success curves as function of step index. Here, step means an iteration of the train-search-select-evaluate cycle. FD refers to adding a neighboring point per each EMT-evaluated data point, where the energy of the second point is estimated with finite-difference method with step length dx (\AA), based on the energy and forces of the EMT point. The notation "exact" refers to evaluating the energy of the second point using EMT.}}\label{cu15_sc}
\end{figure}

In Fig. \ref{cu15_sc}, we show the success curves of different types of global optimization runs for the Cu$_{15}$ cluster in the EMT potential. We perform 40 separate runs with \beacon, L-\beacon, L-\beacon-FD and L-\beacon-exact with different lengths of displacements, where  neighboring training points are included, and 16 separate runs with GOFEE \cite{Malthe2020efficient}. In the figure, the cumulative curves increase by a step each time a single run finds the global minimum with an energy threshold of 0.01 eV/cluster. The threshold value means that we declare a run successful once it hits a true energy that is at maximum 0.01 eV higher than the lowest energy that was found during the runs. The reference energy here corresponds to the geometry of a centered icosahedron of 13 atoms and two adsorbed atoms in neighboring hollow sites of the icosahedron surface, as shown in the inset of Fig. \ref{cu15_sc}. The second lowest-lying local minimum was found at 0.15 eV/cluster above the global minimum, possessing a centered, gyroelongated hexagonal bipyramid. This observation illustrates the ability of the global optimization approach to distinguish between local minima that are close in energy. 

Let us first discuss the three most optimal success curves: \beacon, L-\beacon, and L-\beacon-FD with step length dx=0.001 \AA. The ability of the gradient-trained model to simulate the potential energy surface of EMT is manifested again as 50 \% of the \beacon runs found the true global minimum after only 7 EMT evaluations. The convergence of 7 evaluations would be appealing even for local optimization with the 45 degrees of freedom in the system, although we remind ourselves that convergence here is defined via energy whereas in the context of local relaxation convergence is determined through stricter requirements on the forces in the system. For L-\beacon-FD, the respective number of 50 \% success is 16 evaluations and for L-\beacon where no force information is used, 50 \% success is acquired after 20 evaluations. Our runs with GOFEE \cite{Malthe2020efficient} show that 46 evaluations are required for 50 \% success.  It is worth noting that despite the fast success, the program does not know that it has reached the optimal configuration but keeps searching even after the (known) global minimum is found.

Let us compare our results with random searches, that are shown to be surprisingly efficient when certain chemical intuition is considered \cite{randomsearch}. We run 480 relaxations with the true calculator, EMT, starting from similarly generated random structures as those for the global optimization. The result is that 60, or 12.5 \%, out of all relaxations end up in the global minimum energy structure. If we perform one such EMT relaxation per step, this would statistically result in 61 \% success at step 7, since $\sum_{i=1}^{7} (1-1/8)^{i-1}\times 1/8=0.61$. The important difference is that a single relaxation takes 20-200 EMT calculations, whereas all of our relaxations are performed within the surrogate model and not with the true potential. From this perspective, we conclude again that the model and the global optimization approach of \beacon are together very efficient in the search for the global minimum. Running EMT relaxations is in fact computationally faster than running the relaxations in the surrogate surface within the global search algorithm, as we will discuss further in section \ref{sec:compperf}, but when using the algorithm with DFT this situation is of course completely different. 

\beacon is seen to be the fastest method up to 80 \% success rate.  Most of the L-\beacon-FD (with step length dx=0.001 \AA) runs find the minimum with between 10 and 20 EMT calculations. L-\beacon lags only a little behind, and finally all the three approaches end up with a somewhat similar performance, although the full success of \beacon curve takes slightly more steps with one run finding the correct local minimum after 48 EMT evaluations.

Let us now take a closer look to the different approaches of L-\beacon where the neighboring data points are included into the model, i.e. L-\beacon-FD and L-\beacon-exact. First of all, the success curves with the EMT evaluated neighboring points, that is the L-\beacon-exact curves, are always behind the curve for L-\beacon where no neighboring points are included in the model. With arbitrary increase of step length in L-\beacon-exact, we expect the success curves to reach the performance of L-\beacon at best, since in that case all the data points are more or less individual and the neighboring points do not represent simulating the forces anymore. Also, the step length of dx=0.001 \AA~leads to slightly slower performance than that of dx=0.01 \AA, indicating that the step length of the size 0.001 \AA~is too small to have the desired effect on the model; due to the noise term in the model, the two neighboring points cannot be distinguished properly when they are too close to each other. However, comparing the red and pink curves in Fig. \ref{cu15_sc}b, the smaller step length seems to work similarly between L-\beacon-exact and L-\beacon-FD despite the approximation. This indicates that the linear approximation is accurate enough to produce a reasonable surrogate PES. On the other hand, the step length 0.01 \AA~is clearly too large for a stable model in L-\beacon-FD although it results in better performance with L-\beacon-exact. The value of dx=0.001 \AA~shows the fastest global optimization of all the double-point approaches in numbers of EMT calculations, but the performance is not much better than with L-\beacon. Also, the problem of selecting a suitable step length might be difficult for other systems and more complex potentials. Thus, we do not find inclusion of neighboring points in general beneficial. We will return to this topic later in the case of bulk SiO$_2$.

Comparing our curves with GOFEE, we see that the number of EMT calculations is two-fold or more for GOFEE. As mentioned above, the essential point of interest in the methods is the way in which the energy of the neighboring data point is evaluated. In this respect, L-\beacon-exact is similar to GOFEE. It is observed that L-\beacon-exact is faster in finding the true global minimum, which we attribute to the slight differences in the details of the fingerprint, the kernel function and fitting the hyperparameters along the way. Comparing \beacon and GOFEE, there is about a factor of 6 difference between the number of required energy evaluations.

Finally, let us connect the result with \beacon to the examination of local relaxations above (section \ref{subsec:localrelax}). There, the training set size was 40, and even with 80 relaxations the global minimum was not found. In the global search we find the minimum after only 7 training points. This indicates that updating the model along the run is an important feature of the global optimization algorithm.

\subsection{SiO$_2$}
\begin{figure}[ht]
\includegraphics[width=\figwidth]{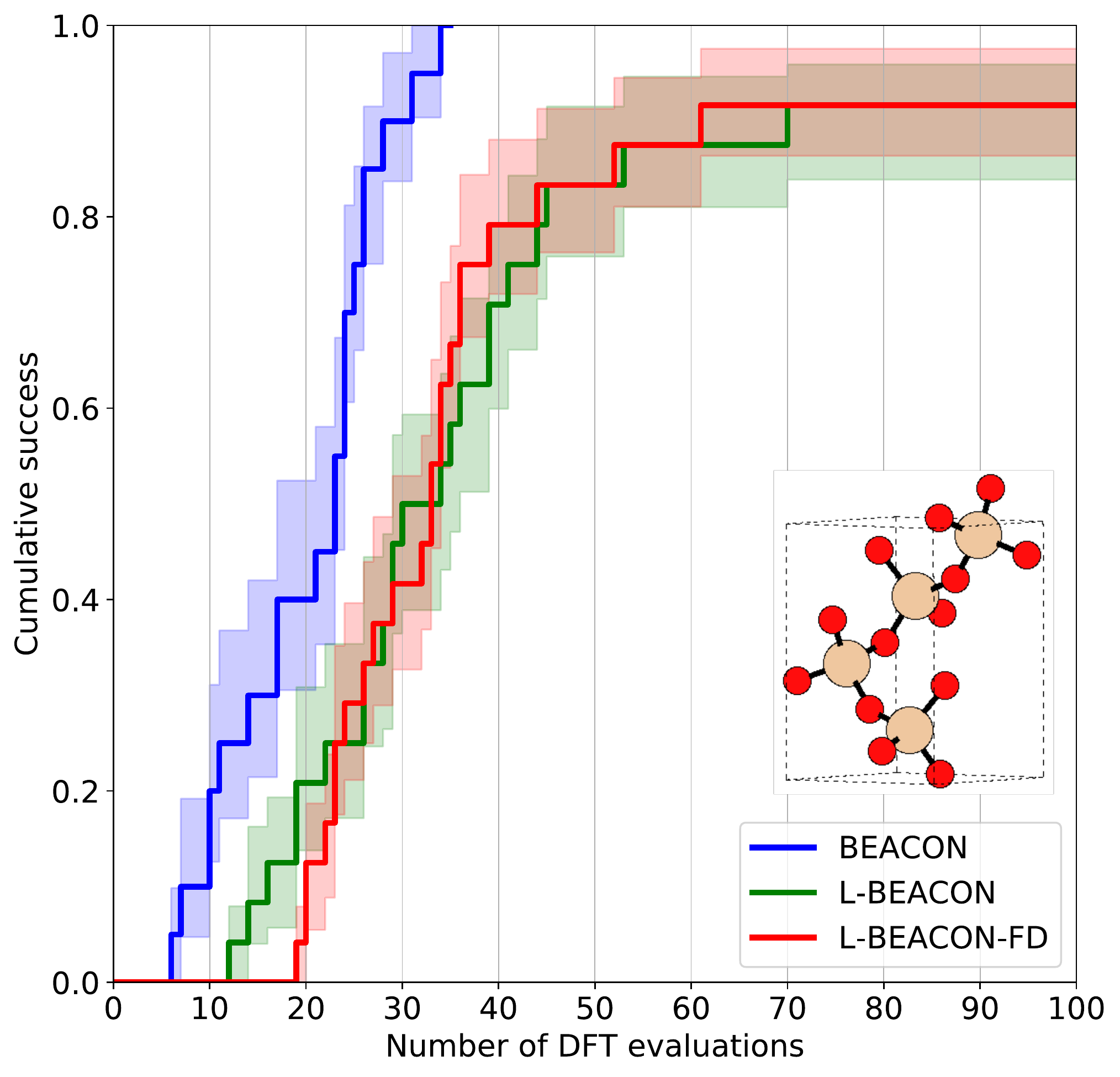}
\caption{\small{SiO$_{2}$ success curves with success threshold of 0.05 eV/cell. The unit cell includes 12 atoms. The global minimum structure is shown in the inset, where 5 extra O atoms are added to the unit cell to illustrate the tetrahedral coordination of Si. For L-\beacon-FD, the step length is 0.001 \AA~in the finite difference method.}}\label{sio2_sc}
\end{figure}

Figure \ref{sio2_sc} shows the success curves for bulk SiO$_{2}$ in the low-cristobalite phase with DFT/PBE. In this case, training the gradients is clearly favourable in the search for the global minimum structure. \beacon finds the correct structure after less than 34 DFT evaluations for all runs. L-\beacon and L-\beacon-FD are slower than \beacon and eventually fail in 2/20 runs to find the true global minimum using 80 DFT calls. 

We saw that for the Cu$_{15}$ cluster the search identified the global minimum based on very few EMT energy and force evaluations compared to what could be expected based on the learning curves. This feature is even more pronounced for the SiO$_2$ system. The learning curves indicate a rather poor accuracy with errors of more than 1 eV/cell (for the 12 atom system) in the range all the way up to 100 training points, but still the global minimum is found within 0.05 eV/cell using only of the order 25 DFT calculations. The explanation for this behavior must have to do with the fact that, in the global search, states around local minima of the PES are of preferential interest and included in the training set. The model is thus exclusively trained to predict a special part of the PES. In the cross validation studies above, the model is trained on a wider range of points and also evaluated broadly in configuration space.

This idea is to some extent illustrated with the simple one-dimensional potential energy surface of the Cu$_3$ cluster discussed above in Fig. \ref{demo}. The evaluation of the model at the minimum point of the well B considerably improves the prediction at the other minimum A, because some of the local bonding characteristics are the same.

Interestingly, the energy-trained models, L-\beacon and L-\beacon-FD, have rather similar performance in the global optimization although the training set of L-\beacon-FD includes twice the number of data points. The difference to the Cu$_{15}$ is the more complex true potential energy surface, introducing more error in the finite-difference method. For the case of SiO$_2$, we assert that a step length of 0.001 \AA~is too small for the energies of the neighboring points to be distinguishable by the Gaussian process, resulting in failure of simulating the slopes of the PES. Increasing the step length increases the risk of running into problems with an unstable Gaussian process, as observed with Cu$_{15}$. We thus conclude, somewhat at variance with Bisbo and Hammer \cite{Bisbo:2020vo}, that adding neighboring data points, as done in L-\beacon-FD, L-\beacon-exact, or GOFEE, is not beneficial in general for the Bayesian approach to global optimization.

Let us now investigate how the success curves depend on the threshold value for the energy. In Fig. \ref{sio2_sc} an energy threshold of 0.05 eV/cell is used for SiO$_2$, and we show the success curves with energy thresholds of 0.2 eV/cell and 0.01 eV/cell in the Supplemental material \cite{suppinfo}. We see that \beacon is more efficient than the L-\beacon methods, which are quite similar to each other, and therefore the overall analysis is not too sensitive to the choice of the threshold value. However, the choice of an appropriate threshold may depend on the system being investigated. For example, an energy threshold of 0.2 eV/cell for the Cu$_{15}$ cluster has the consequence that finding the second lowest local minimum is also counted as a successful  run. For SiO$_2$, we did not determine the second most stable structure explicitly, but no other stable structure was found within the highest threshold of 0.2 eV/cell. On the other hand, small thresholds like that of 0.01 eV/cell might fall below the accuracy of the DFT implementation, parameters in use (such as k-point density), convergence criteria of the self-consistent cycle etc., making the judgment whether the global minimum was found or not. The threshold of 0.05 eV/cell seems to be a good compromise for global optimization of systems with 10-20 atoms using DFT calculators like GPAW with default settings. In this work, we use the threshold value of 0.05 eV/cell for all systems studied with DFT.

\begin{figure}
    \centering
    \includegraphics[width=\figwidth]{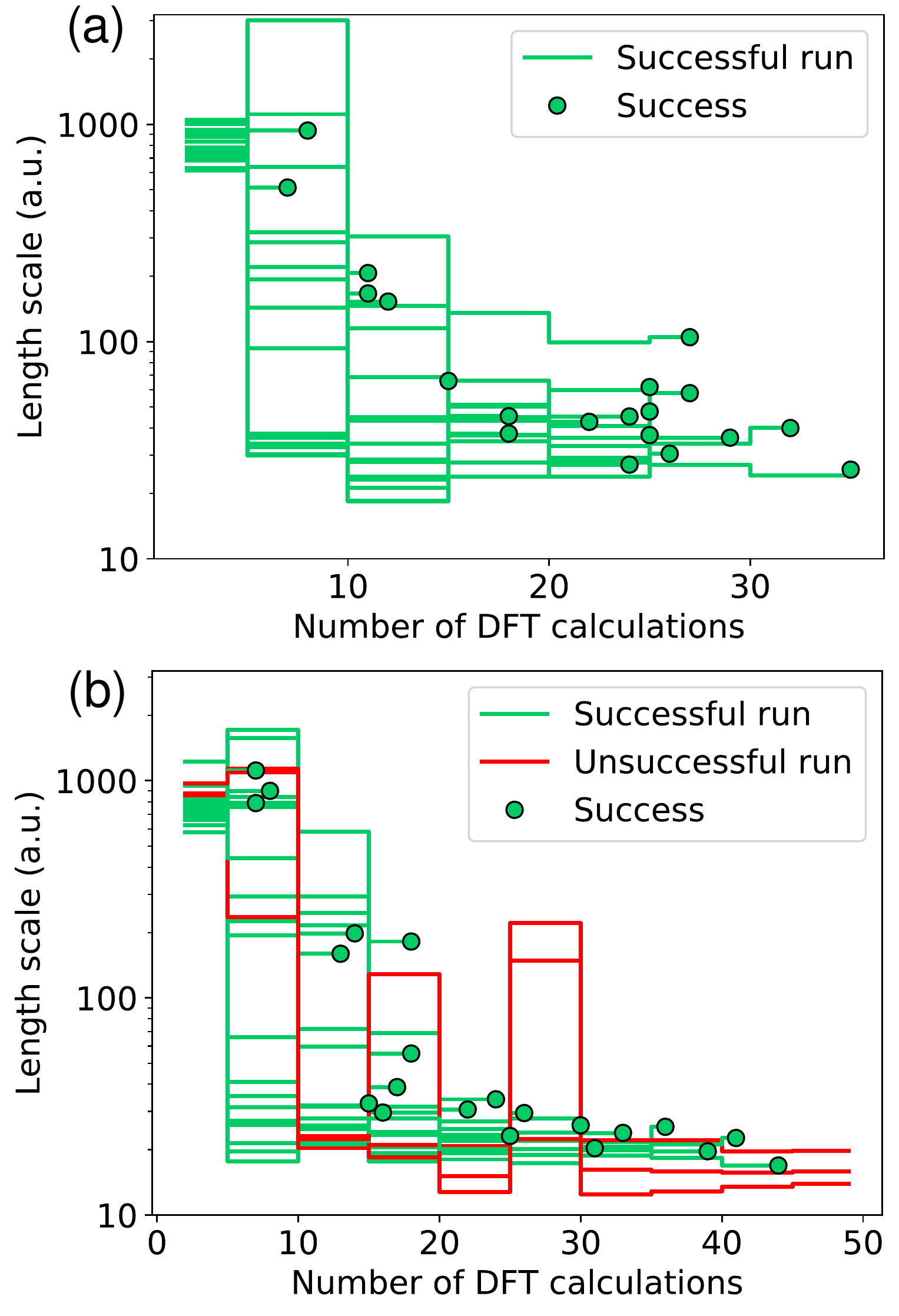}
    \caption{Evolution of the updated length scales of the Gaussian process kernel during the runs of SiO$_2$ in (a) \beacon and (b) \beacon without lower bound for the updated length scales.}
    \label{sio2_scales}
\end{figure}

The length scale is updated every 5 steps by maximizing the marginal log-likelihood as discussed above in the section \ref{sec:glopt}. Furthermore, as also discussed above, the length scale is bounded from below by the mean value of all distances between the training data points in fingerprint space. In Fig. \ref{sio2_scales}, we show the evolution of the fitted length scale in the global optimization runs of SiO$_2$ when constraining the fitted length scales from below and when not. For both cases, we observe that after the first fitting at 5 DFT calculations there is a huge variance in the optimal length scales over the different runs: the values range from 30 to 3000. As the searches proceed, this range gets narrower, and after 20 DFT calculations the updated scales vary only between 25 and 100 when constraints are turned on. The corresponding figure for L-\beacon runs is shown in Supplemental material \cite{suppinfo}. For L-\beacon, most of the values lie below 100, but the updating also converges to higher values along the search. This does not seem to be a problem though, as global minima are also found with the large scales.

For \beacon, the length scales where the correct global minimum structure is found are gradually decreasing as the search proceeds. Furthermore, it can be seen that if the length scale is below 100, the global minimum is never found in fewer than 15 steps. Apparently, the length scales obtained by maximizing the marginal log-likelihood are not necessarily optimal in the early part of the global search, where exploration is particularly important. The large variation in the length scales in the beginning of the search is hardly surprising in light of the small number of training points. In this perspective, it seems reasonable to limit the acceptable length scales from below to prevent strong overfitting in the beginning of the search. It appears from the data in Fig. \ref{sio2_scales} that the lower bound could be set even higher in the beginning of the runs.

Without the lower bound on the length scale, three of the runs fail to find the global minimum in 50 DFT calculations as shown in Fig. \ref{sio2_scales}b. It is clear that the failing runs after 20 steps have very short length scales and even though the length scale is increased after 25 steps they fall back to a less exploratory mode, where the global minimum is not identified.

\subsection{[Ta$_2$O$_5$]$_x$}
Ta$_2$O$_5$ is an optically interesting material, whose crystal structure is still under debate \cite{PerezWalton2015, Yang2018, Yuan2019}. Furthermore, clusters of the material might be of interest in photocatalysis \cite{Srivastava:2017ko, Chen:2003dh}.

\begin{figure}
    \centering
    \includegraphics[width=\figwidth]{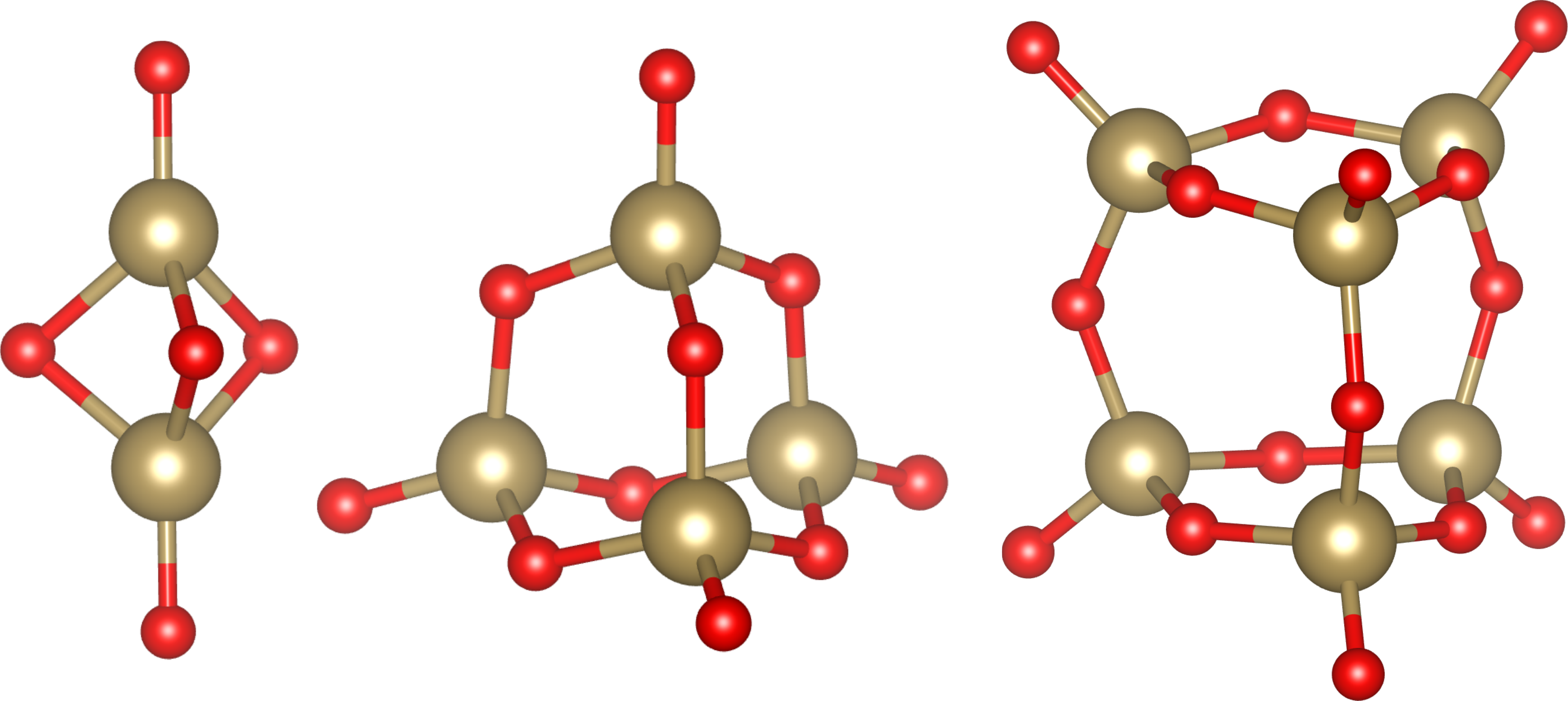}
    \caption{Global minimum structures of Ta$_2$O$_5$  (D$_\mathrm{3h}$ symmetry), Ta$_4$O$_{10}$ (T$_\mathrm d$) and Ta$_6$O$_{15}$ (D$_\mathrm{3h}$) clusters as found by \beacon.}
    \label{fig:ta2o5}
\end{figure}

To test our approach, we investigate small clusters with the composition [Ta$_2$O$_5$]$_x$ with $x=1,2,3$ corresponding to the stoichiometry of the bulk material.
The structures that we find to minimize the potential energy are shown in Fig. \ref{fig:ta2o5}. All the globally optimal structures for clusters are such that O is sticking out of the TaO core, making it impossible to use these units as building blocks for a bulk system while preserving the correct stoichiometry Ta$_2$O$_5$. Nevertheless, it is interesting that in all the structures every Ta atom has a similar bonding environment to four O atoms, one of which is pointing outwards from the cluster. Interpreting this as a double bond makes the oxidation number of each Ta atom to be +5.

In any case, the global optimization of the clusters provides a good demonstration of our method. Of course there is no rigorous proof that the obtained structures are in fact the global minimum energy structures, but some indications of this are obtained by just repeating the searches and noting the variety of structures visited during the search.
The shown structures were found in 4/4 runs for Ta$_2$O$_5$, 5/6 runs for [Ta$_2$O$_5$]$_2$, and 4/8 runs for [Ta$_2$O$_5$]$_3$. The second lowest local minimum for Ta$_2$O$_5$ was observed at 0.16 eV/cluster higher than the lowest one, and this was visited in all runs. This again indicates that the model and the acquisition function are able to identify small energy differences between different structures.

The average number of required single-point DFT calculations before hitting the global minimum structure (with threshold of 0.05 eV/cluster) was 42 for Ta$_2$O$_5$, 40 for [Ta$_2$O$_5$]$_2$, and 35 for [Ta$_2$O$_5$]$_3$, calculated among the successful runs. The small number of steps required is a highly desired result, because it means that one can limit the length of the \beacon runs. This does not only have the advantage that the total number of DFT calculations is small, but long runs of \beacon with many steps lead to surrogate models with many data points, which require more memory and computational time. It is interesting that the average number of DFT calculations is smallest for the largest cluster where the number of degrees of freedom is the largest. This could be explained by the fact that whenever we train the model with a larger cluster, more training data is available since the number of trained gradients is larger. However, we note that in general larger systems can be expected to exhibit considerably more local minima to be explored making the global search more difficult.

\begin{figure*}[ht]
\includegraphics[width=\widefig]{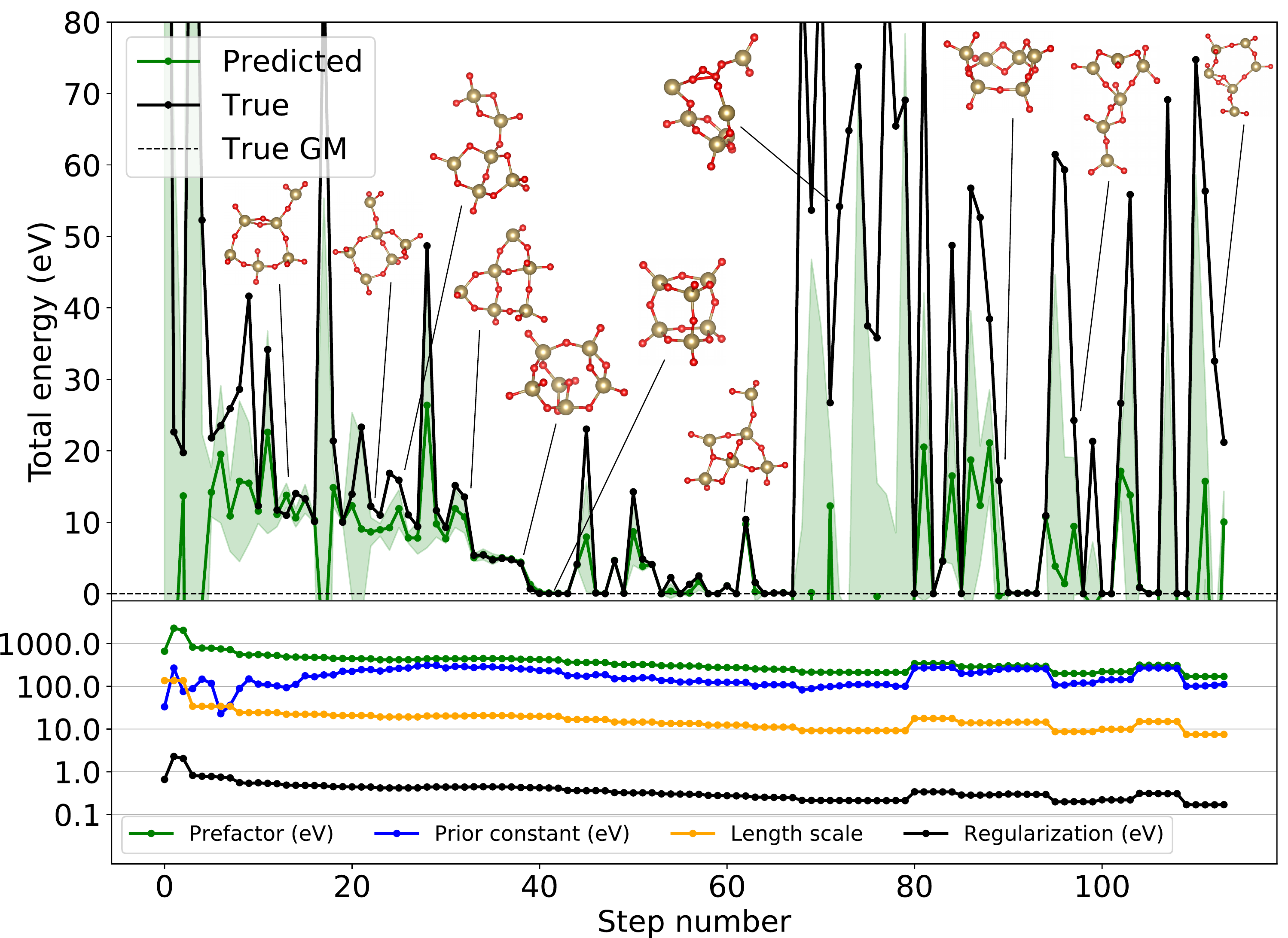}
\caption{\small{Energy, prediction, and hyperparameter evolution in a single global optimization run for a Ta$_6$O$_{15}$ cluster. Also, selected structures are shown that were evaluated with DFT along the run. The global minimum structure is visited at step number 42, and at various steps after that. The green area shows the estimated uncertainty of the prediction. The prior constant is shown with respect to the global minimum energy.}}\label{ta6o15_run}
\end{figure*}

We now take a closer look at some of the \beacon runs in order to get a better understanding of how the algorithm behaves. We first investigate why some of the runs fail for the larger clusters by checking how the global minimum structure is predicted with surrogate models of the unsuccessful runs. That is, we use the surrogate models at step number 40 of the global optimization runs (42 training points), and perform a local relaxation on this surrogate potential energy surface starting from the global minimum structure. In every case, the relaxation does not change the structure significantly, and the acquisition function of the relaxed structure is low enough so that it would have been selected for DFT evaluation in the original \beacon run. We thus conclude that for these runs, the search within the surrogate space is insufficient whereas the accuracy of the model is good enough to find the global minimum.

In Fig. \ref{ta6o15_run}, we show how a single run of Ta$_6$O$_{15}$ proceeds along the search. The search starts with high-energy structures where the prediction and the true energy do not match at all, but as the high uncertainties show, the model knows it might be wrong about the predictions. As the search continues, different structures are suggested and evaluated with a good agreement between the predictions and the true energies, taking the estimated uncertainties into account. The global minimum structure is found at step 42 in this particular run, and the exploration continues after that, as indicated by the higher-energy structures that are visited between step 42 and step 68. As noted in the description of the algorithm, we remove structures, which have been obtained by local relaxations in the surrogate model, if a bond distance is smaller than 0.7 times the sum of the covalent radii of the atoms in the bond. What happens at step 68 in this \beacon run is that the model begins to develop local minima with O-O bond lengths, which are just above this threshold. So the new predicted structures contain unphysically short O-O bonds or even clustering oxygens linked to the rest of the cluster. The model predicts their energies to be very low, and they are therefore always selected by the acquisition function (eq. \ref{acqf}). However, after step 80 we see that the model has learned that the unphysical structures have high energies, and the search continues in a more reasonable way both exploring new areas and exploiting the known structures, the global minimum included.

This issue illustrates how additional conditions on the search might be helpful to avoid unphysical structures, but also that the surrogate model might react to these in unexpected ways.

Let us briefly look at how the hyperparameters evolve as shown in Fig. \ref{ta6o15_run}. To first recapitulate, we have four hyperparameters: the length scale in the kernel, the kernel prefactor, the prior constant, and the noise parameter. For simplicity, we keep the ratio between the noise and the prefactor fixed, because the maximization of the log-likelihood can then be done analytically with respect to both the prefactor and the prior constant. Only the length scale has to be obtained numerically and this update is done every 5 steps. We see in Fig. \ref{ta6o15_run} that the prior constant has some variation in the beginning but the changes become smoother as more training data is acquired. The dramatic changes in the model at step 68 also lead to a recalibration of the prior constant. It should be noted that the prior constant is not only a weighted mean of the observed energies, because the gradients also play a role in the determination as seen from Equation (\ref{prior_constant}).

The length scale is significantly reduced during the run, and we take this as an indication that the model is initially focusing on getting the large-scale features of the PES correct with subsequent refinements. This is an appropriate behavior for a global search strategy that we already examined in the case of SiO$_2$. 

The kernel prefactor also gradually decreases during the global optimization with the most significant changes every 5 steps when the length scale is updated. The prefactor is unimportant for the prediction of energies and forces as can be seen from Equation~(\ref{gp_mean}), but it plays a major role for the estimated uncertainties. The decaying value might therefore indicate an increasing confidence of the model. The reduction of the prefactor can also be seen as coupled to the change in the length scale. The uncertainty at a particular point in fingerprint space is roughly given by the kernel function to neighboring data points. This estimate involves both the prefactor of the kernel function and the distances to the neighboring data points measured in units of the length scale. A reduction of the length scale thus leads to a less "stiff" model with larger variances and this is to some extent compensated by the reduced prefactor.
The regularization follows exactly the variation of the kernel prefactor because the quotient of these quantities is kept constant throughout the run. 

In the Supplemental material \cite{suppinfo}, we show another, similar run to that in Fig. \ref{ta6o15_run}. Although the global minimum energy structure is not found, the overall behaviour of the predictions and hyperparameters is similar with a good balance between exploration and exploitation. Even the stage of unphysical structures with short O-O bonds is the same after 60 steps, after which the search becomes stable again at around step 80. Although the prior constant goes below the global minimum energy in the beginning, we note that it does not have a negative effect on the model or the search: in the beginning the length scale is relatively long, and therefore all the points in the (reasonable) coordinate space are considered close to each other compared to the length scale, and consequently the model is not very sensitive to the absolute value of the prior.

An important feature observed in both runs is that the estimated uncertainties are reasonable so that when the error of the prediction is large, the model knows that it might be wrong. This is an essential property for using the lower confidence bound as the acquisition function, Equation \ref{acqf}, to control the balance between exploration and exploitation in the global search.

\subsection{ZrN-O surface}
\begin{figure}[ht]
\includegraphics[width=\figwidth]{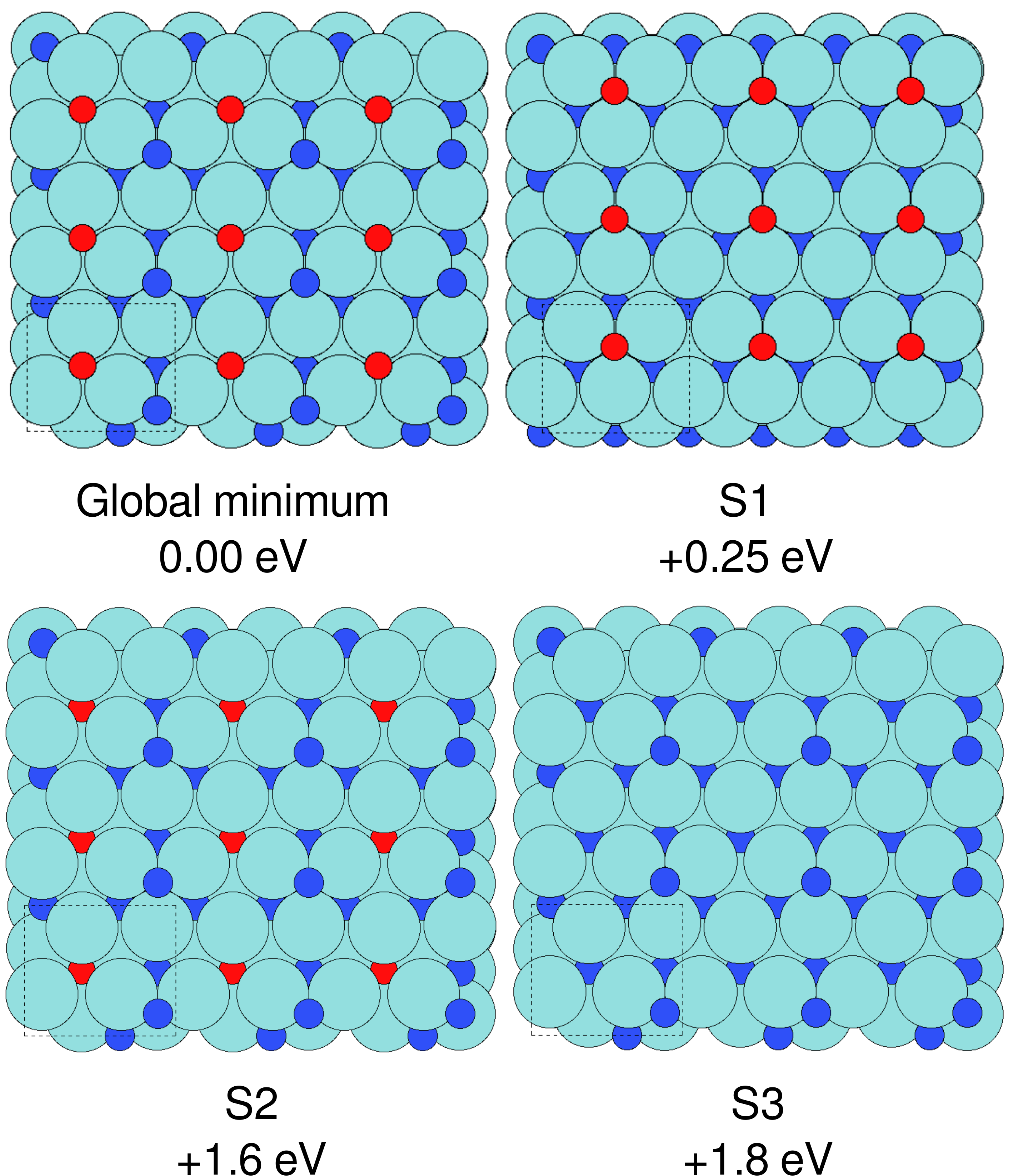}
\caption{\small{Investigated structures for ZrN-O surface and their respective total energies. The structures are shown from above the surface. See text for details about the structures. Light blue: Zr, sky blue: N, red: O.}}\label{zrn_structures}
\end{figure}

As the last demonstration, we briefly illustrate the applicability of the approach to surface structures. Recently, a high catalytic activity of ZrN for oxygen reduction was observed \cite{Yuan2020}. To see whether anything interesting occurs on the ZrN surface, exposed to oxygen, we investigate the surface structure of ZrN with adsorbed oxygen using the global optimization method. We use a surface slab of 4 layers with the 2 bottom layers fixed during the optimization. The unit cell is orthogonal containing 16 zirconium atoms, 16 nitrogen atoms and 1 oxygen atom corresponding to a coverage of one oxygen atom per four zirconium atoms in the surface layer. The electronic exchange-correlation effects are modelled using RPBE \cite{RPBE}. With this setup, the global optimization algorithm finds that a structure where the oxygen atom and also one of the nitrogen atoms occupy the hollow surface sites minimizes the potential energy of the system as shown in Fig. \ref{zrn_structures}. Consequently, there is a nitrogen vacancy in the first layer, below the oxygen atom. The Zr lattice stays close to the cubic (111) surface form, although the 3 Zr atoms that are neighbors to oxygen tend to move away from the oxygen atom and lie closer to the nitrogen on the surface, as compared to the bulk structure.

To verify the surprising finding that one of the N atoms in the unit cell prefers a surface site, we relax the obtained global minimum structure with a maximum residual force of 0.05 eV/\AA. In addition, we relax three other structures that are built manually (see Fig. \ref{zrn_structures}). S1: This is a structure where only oxygen is on the surface and the ZrN lattice has its bulk form. S2: In this structure a single nitrogen atom is on the surface and oxygen is moved to the vacancy left behind by the nitrogen. S3: A structure where a single nitrogen is on the surface and oxygen is in the second N-layer. Comparing with these structures, the structure identified by the global search has the lowest energy with energy differences of 0.25 eV/cell, 1.6 eV/cell and 1.8 eV/cell to S1, S2, and S3, respectively. The method thus finds a local minimum structure that is lower in energy than the most intuitive configurations.

\begin{figure*}[ht]
    \centering
    \includegraphics[width=\widefig]{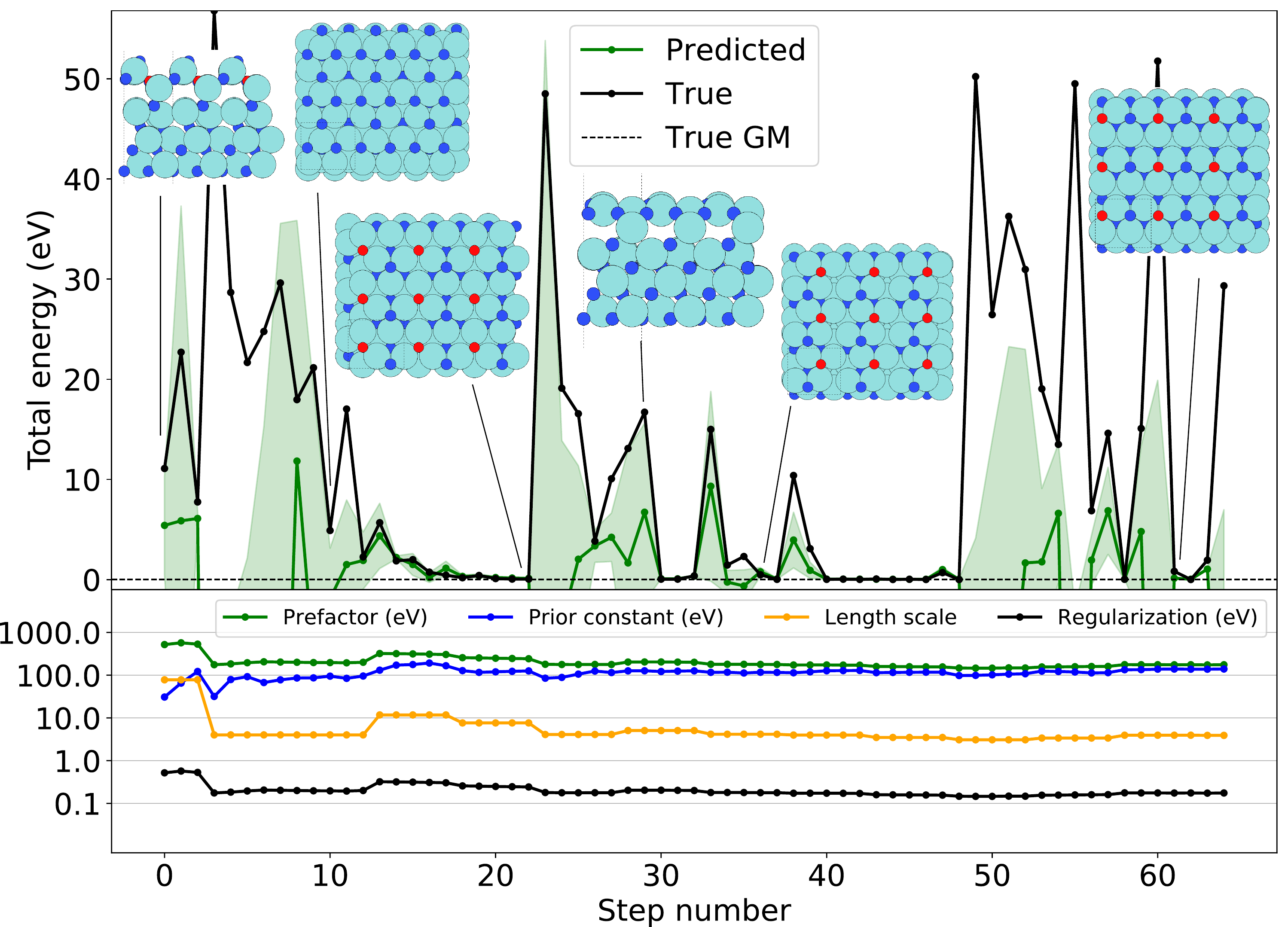}
    \caption{A single global optimization run of ZrN-O surface. The global minimum structure was found at step 22; the structure is shown in the figure.}
    \label{zrno_single}
\end{figure*}

The progress of one of the successful global optimization runs is shown in Fig. \ref{zrno_single}. It illustrates that the method is visiting a diverse set of structures. For the first 11 steps, the program produces rather unrealistic structures with energies above 10 eV/cell higher than the global minimum. After step 11, the low-energy structures are exploited more thoroughly, and the global minimum structure is found at step number 22. The program continues with a balance between exploring structures at fairly high energies and identifying competing low energy structures, for example at steps 36 and 61. It is also notable that some non-trivial Zr surface structures are explored, such as those at steps 1 and 29. In general, we note that a fair share of the structures investigated are pretty high in energy. This seems to be necessary to achieve a proper exploration and training of the model.

\subsection{Other systems}
\begin{figure*}
    \centering
    \includegraphics[width=\widefig]{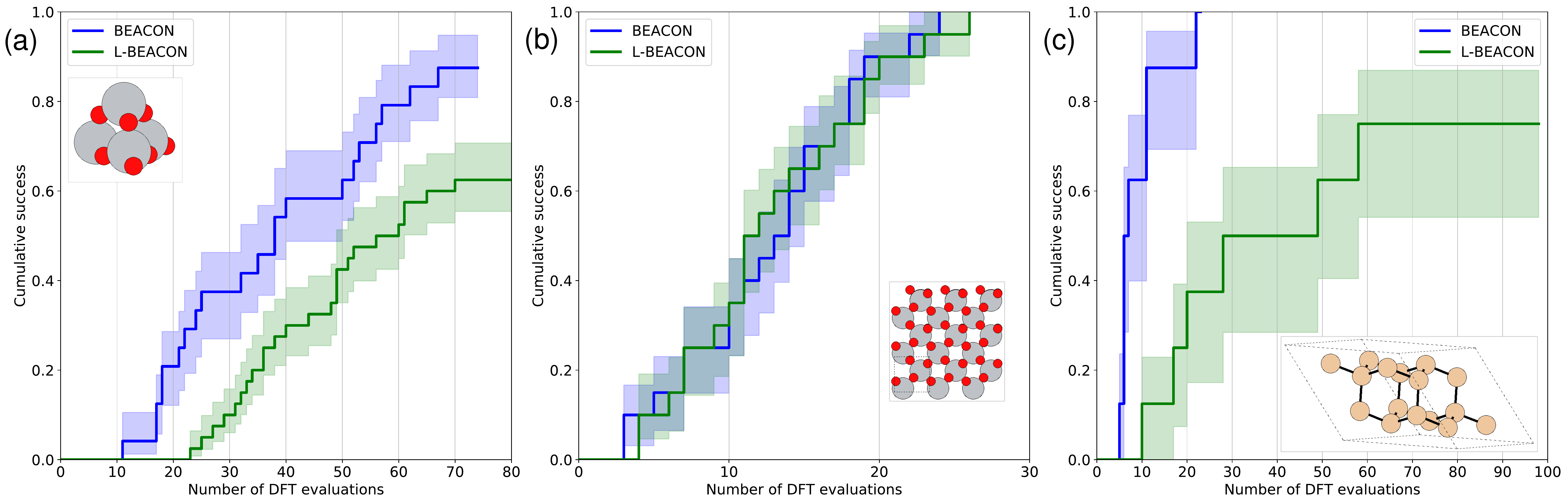}
    \caption{(a) Ti$_4$O$_8$ cluster, (b) bulk TiO$_2$ and (c) bulk Si/diamond success curves with success threshold 0.05 eV/cell. For Ti$4$O$_8$ cluster, tight-binding DFT was used as true potential \cite{DFTBplus, DFTBparams}, whereas DFT was used for the bulk systems with the same setups as for the SiO$_2$. The unit cells of TiO$_2$ and Si include 12 and 16 atoms, respectively.}
    \label{othersystems}
\end{figure*}

One of the main points of this paper is to show how including gradients in the Gaussian process affects the performance of Bayesian global optimization in comparison with GOFEE \cite{Malthe2020efficient} where only energies are used for training. Our results with both learning curves and success curves indicate that adding the gradient information into the model improves the performance of the search. In Fig. \ref{othersystems}, we show success curves for three more systems, to investigate the effect of training on the gradients as well as the energies. We run several optimizations for a Ti$_4$O$_8$ cluster, bulk TiO$_2$, and bulk silicon. The bulk TiO$_2$ system consists of 12 atoms and the unit cell is fixed as appropriate for the rutile phase. The bulk silicon system consists of 16 atoms and the unit cell is fixed corresponding to the diamond lattice. For the Ti$_4$O$_8$ cluster and bulk silicon the improvement is notable. In contrast, and a bit surprisingly, training on the gradients does not have any effect on the overall performance for bulk TiO$_2$, so the potential gain of including gradients does depend on the system under study. However, at this point, we cannot tell how much different properties like size, symmetry, number of elements, shape of the true potential etc. matter for the acceleration obtained by using gradients.

\section{Computational time and memory limitations}\label{sec:compperf}
The primary goal for this work is to reduce the number of expensive DFT calculations necessary to find the global minimum of a PES. However, it should be noted that in some cases the processor time needed to run the relaxations on the surrogate surface may become comparable to the time spent on DFT, especially if the DFT implementation is parallelized efficiently while the Gaussian process is not. In our current implementation each surrogate relaxation is always run on a separate, single processor, but no further parallelization is performed. In the beginning, most of the time of the surrogate model is spent on calculating the fingerprints and their gradients at every step of the local relaxations. Later in a global optimization run, as the training set size becomes larger, calculating the Hessians of the kernel function in fingerprint space is the computational bottleneck. (See the computational times of training and predicting in Supplemental material \cite{suppinfo}.) In our examples, training the model with gradients typically takes more than one order of magnitude more time compared to using the energies alone. Time consumed in predicting is roughly the same with few training points, but as the number increases, ever larger Hessian matrices need to be calculated with the gradients, leading to an increase in computer time. Predicting without training the gradients is faster since the kernel Hessians are not calculated, and the linear algebra is applied to much smaller matrices. The numerical updating of the length scale involves training the model at each optimization step, and with a large number of atoms and training set sizes, this becomes too heavy in practice with the gradient-trained models unless the training is parallelized. Eventually, if the DFT calculation is fast enough, the energy-trained L-\beacon might become faster in total processor time than \beacon even if more DFT calculations are required to train a sufficient model. However, as we observe from the success curves, \beacon is more robust in finding the global minimum and does not get stuck as often as L-\beacon.

The memory usage is limited by inversion of the C-matrix, as required by equation \ref{gp_mean}, that scales as $\mathcal{O}(n^2)$ memory-wise \cite{GPMin, koinstinen2017} where $n$ is the order of the square matrix, i.e. in this context $n=N(1 + 3N_{\mathrm{atoms}})$. If memory or speed becomes an issue with large systems, one is forced to use L-\beacon (or switch to L-\beacon on the fly during the optimization), where we have merely $n=\mathrm{number~}\mathrm{of~}\mathrm{training~} \mathrm{points}$.

\section{Conclusions}\label{sec:concl}
We think that the use of Bayesian strategies and more specifically Gaussian processes for global atomic structure determination is only at its beginning. As demonstrated by Bisbo and Hammer \cite{Malthe2020efficient, Bisbo:2020vo}, a surrogate model based on a global atomic fingerprint in combination with a Bayesian optimization can outperform earlier global optimization methods by orders of magnitude in reduced computer power. In the present paper we show that including gradient information, {\em{i.e.}} the atomic forces, in the training of the model in most cases leads to a further reduction in the number of DFT calculations necessary to identify the global minimum energy structure. The approach was successfully applied to clusters, surfaces, and bulk systems.

However, many aspects of the approach are still unexplored. The surrogate models are not particularly accurate as shown by the learning curves, but still they work very well in the optimization process. So far only a single global fingerprint was investigated, and it is not known if other fingerprints like SOAP \cite{SOAP}, MBTR \cite{MBTR}, or FCHL \cite{FCHL} would perform even better. The way of suggesting new candidate structures could potentially also be improved, for example in combination with a genetic algorithm, and other acquisition functions may be relevant. Finally, the approach could be combined with other artificial intelligence techniques providing additional guiding of the search.

The code for \beacon is available at \url{https://gitlab.com/gpatom/ase-gpatom}. It is integrated with the Atomic Simulation Environment (ASE) \cite{ase-paper, ase}, so that any energy and force calculator supported by ASE can be used together with \beacon. In the present implementation the unit cell is kept fixed during the search. However, it should be possible to update the unit cell based on the currently applied fingerprint, and we expect to implement that in the near future.

\begin{acknowledgments}
We acknowledge support from the VILLUM Center for
Science of Sustainable Fuels and Chemicals, which is funded by
the VILLUM Fonden research grant (9455).
\end{acknowledgments}

\bibliography{bibliography}

\end{document}


\centering
{\large\bf{Supplemental Material for:\\ Global optimization of atomic structures with gradient-enhanced Gaussian process regression}}\\[10pt]

Sami Kaappa, Estefan\'ia Garijo del R\'io, Karsten Wedel Jacobsen\\

{\em{Department of Physics, Technical University of Denmark, Kongens Lyngby, Denmark}}\\

(Dated: \today)\\[30pt]
\flushleft
{\bf{Formulas for kernel and fingerprint gradients}}\\[15pt]
The squared exponential kernel is of Gaussian shape
\begin{equation}
k(\rho_1,\rho_2) = \sigma^2\exp\left(\frac{-D(\rho_1, \rho_2)^2}{2l^2}\right)
\end{equation}
where $D$ stands for a specific distance function between the arguments. In this work, the distance is chosen to be the Euclidean distance
\begin{equation}
    D(\rho_1, \rho_2) = \left|\rho_1 - \rho_2 \right| = \left[\sum_i (\rho_{1i} - \rho_{2i})^2\right]^{1/2}
\end{equation}
\\
In the following, let's denote $k = k(\rho_1,\rho_2)$ and $D = D(\rho_1,\rho_2)$. Let us differentiate the kernel function with respect to a single Cartesian coordinate of an atom, $\mathbf r_n$, that is the coordinate with index $n$ in the set of atoms corresponding to the fingerprint $\rho_1$:
\begin{equation}\label{dkx}
\nabla_{1,n}k = \frac{\partial k}{\partial \mathbf r_n} = \frac{\partial k}{\partial D} 
\frac{\partial D}{\partial \mathbf r_n}
\end{equation}
The first factor is straightforwardly
\begin{equation}
\frac{\partial k}{\partial D} = -\sigma^2\frac{D}{l^2}
    \exp\left(-D^2/2l^2\right).
\end{equation}
The second factor becomes
\begin{equation}\label{dDdrn}
\frac{\partial D}{\partial \mathbf r_n} = \frac{1}{D} (\rho_1 - \rho_2)
\cdot
\frac{\partial \rho_1}{\partial \mathbf r_n}.
\end{equation}
The fingerprint gradient is also straightforward to calculate. The radial part becomes
\begin{align}
\nonumber \frac{\partial \rho_{AB}^R}{\partial \mathbf r_n} = \sum_{\substack{i\in A \\ j\in B}} & \delta_{in}\left(\frac{1}{r_{nj}^2}f_c(r_{nj})\frac{r-r_{nj}}{\delta_R^2} - \frac{2}{r_{nj}^3} + \frac{2}{R_c^3} \right)\frac{1}{r_{nj}} \exp\left(\frac{-|r-r_{nj}|^2}{2\delta_R^2}\right)\left( \mathbf r_n - \mathbf r_j\right)\\
 + & \delta_{jn}\left(\frac{1}{r_{in}^2}f_c(r_{in})\frac{r-r_{in}}{\delta_R^2} - \frac{2}{r_{in}^3} + \frac{2}{R_c^3} \right)\frac{1}{r_{in}} \exp\left(\frac{-|r-r_{in}|^2}{2\delta_R^2}\right)\left( \mathbf r_n - \mathbf r_i\right)
\end{align}
with the notation matching with the main text. The angular part has
\begin{align}
\frac{\partial \rho_{ABC}^\alpha}{\partial \mathbf r_n} = \sum_{\substack{i\in A \\ j\in B \\ k\in C}} & \frac{\partial f_c(r_{ij})}{\partial \mathbf r_n}f_c(r_{jk})\exp(-|\theta - \theta_{ijk}|^2/2\delta_\alpha^2)\\
\nonumber + & f_c(r_{ij})\frac{\partial f_c(r_{jk})}{\partial \mathbf r_n}\exp(-|\theta - \theta_{ijk}|^2/2\delta_\alpha^2) \\
\nonumber + & f_c(r_{ij}) f_c(r_{jk}) \frac{\partial }{\partial \mathbf r_n}\exp(-|\theta - \theta_{ijk}|^2/2\delta_\alpha^2)
\end{align}
where
\begin{equation}
    \frac{\partial f_c(r_{ij})}{\partial \mathbf r_n} = \frac{6}{R^2}\left(\frac{r_{ij}}{R} - 1\right)\left(\delta_{ni}(\mathbf r_{n} - \mathbf r_j) + \delta_{nj}(\mathbf r_{n} - \mathbf r_i)\right)
\end{equation}
and
\begin{equation}
    \frac{\partial }{\partial \mathbf r_n}\exp(-|\theta - \theta_{ijk}|^2/2\delta_\alpha^2) = \frac{\theta - \theta_{ijk}}{\delta_\alpha^2}\exp(-|\theta - \theta_{ijk}|^2/2\delta_\alpha^2) \frac{\partial \theta_{ijk}}{\partial \mathbf r_n}.
\end{equation}
Since the angle between the atoms is
\begin{equation}
    \theta_{ijk} = \arccos\left( \frac{\mathbf r_{ij} \cdot \mathbf r_{jk}}{r_{ij}r_{jk}}\right),
\end{equation}
the gradients of the angles are given by
\begin{align}
    \frac{\partial \theta_{ijk}}{\partial \mathbf r_n} = -\frac{1}{\sqrt{1-p^2}}\Bigg( & \delta_{in}\left(\frac{-\mathbf r_{jk}}{r_{ij}r_{jk}} + p\frac{\mathbf r_{ij}}{r_{ij}^2} \right) \\
    \nonumber + & \delta_{jn}\left(\frac{\mathbf r_{jk} - \mathbf r_{ij}}{r_{ij}r_{jk}} + p\left(\frac{\mathbf r_{jk}}{r_{jk}^2} - \frac{\mathbf r_{ij}}{r_{ij}^2}\right)\right)\\
    \nonumber + & \delta_{kn} \left(\frac{\mathbf r_{ij}}{r_{ij}r_{jk}} - p\frac{\mathbf r_{jk}}{r_{jk}^2} \right)\Bigg)
\end{align}
with
\begin{equation}
    p = \frac{\mathbf r_{ij} \cdot \mathbf r_{jk}}{r_{ij}r_{jk}}.
\end{equation}
The so-called Hessian of the kernel, that appears in Eq. 2 of the main text, becomes
\begin{align}
    \nabla_{2,m}\left(\nabla_{1,n}k\right) &= \frac{\partial}{\partial \mathbf r_m} \left( \frac{\partial k}{\partial \mathbf r_n}\right)\\
    \nonumber &= \frac{1}{l^2}k\Bigg( \frac{D^2}{l^2}\frac{\partial D}{\partial \mathbf r_m} \frac{\partial D}{\partial \mathbf r_n}
    + \frac{\partial \rho_1}{\partial \mathbf r_n} \cdot \frac{\partial \rho_2 }{\partial \mathbf r_m}\Bigg)
\end{align}
for which all the ingredients are given above. Here the derivative for $\mathbf r_m$ is for a coordinate with index $m$ in $\rho_2$, the second argument of $k$.\\[20pt]

{\bf{Force prediction in energy-trained model}}\\[15pt]
In the model that is trained on energies only, the forces on the surrogate potential energy surface have simply
\begin{equation}
    -F_n(x) = \frac{\partial E(\mathbf x)}{\partial \mathbf r_n} = \frac{\partial E_p(\mathbf x)}{\partial \mathbf r_n} + \frac{\partial K(\rho(\mathbf x), P)}{\partial \mathbf r_n}C(P, P)^{-1} (y-E_p(X)),
\end{equation}
where $E_p(\mathbf x)$ is the prior function for the energy, and the matrix $\frac{\partial K(\rho(\mathbf x), P)}{\partial \mathbf r_n}$ consists of the gradients of the kernel functions with respect to each atomic coordinate of the system.\\[20pt]

{\bf{Supplementary results}}

~\\[20pt]
\begin{figure}[h!]
    \centering
    \includegraphics[width=0.24\linewidth]{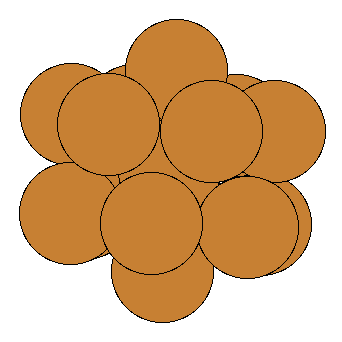}
    \caption{The second global minimum structure of Cu$_{15}$ in EMT potential. The structure is a centered, gyroelongated hexagonal bipyramid, and belongs to the point group D6d.}
    \label{sio2_thresholds}
\end{figure}

\begin{figure}[h!]
    \centering
    \includegraphics[width=\linewidth]{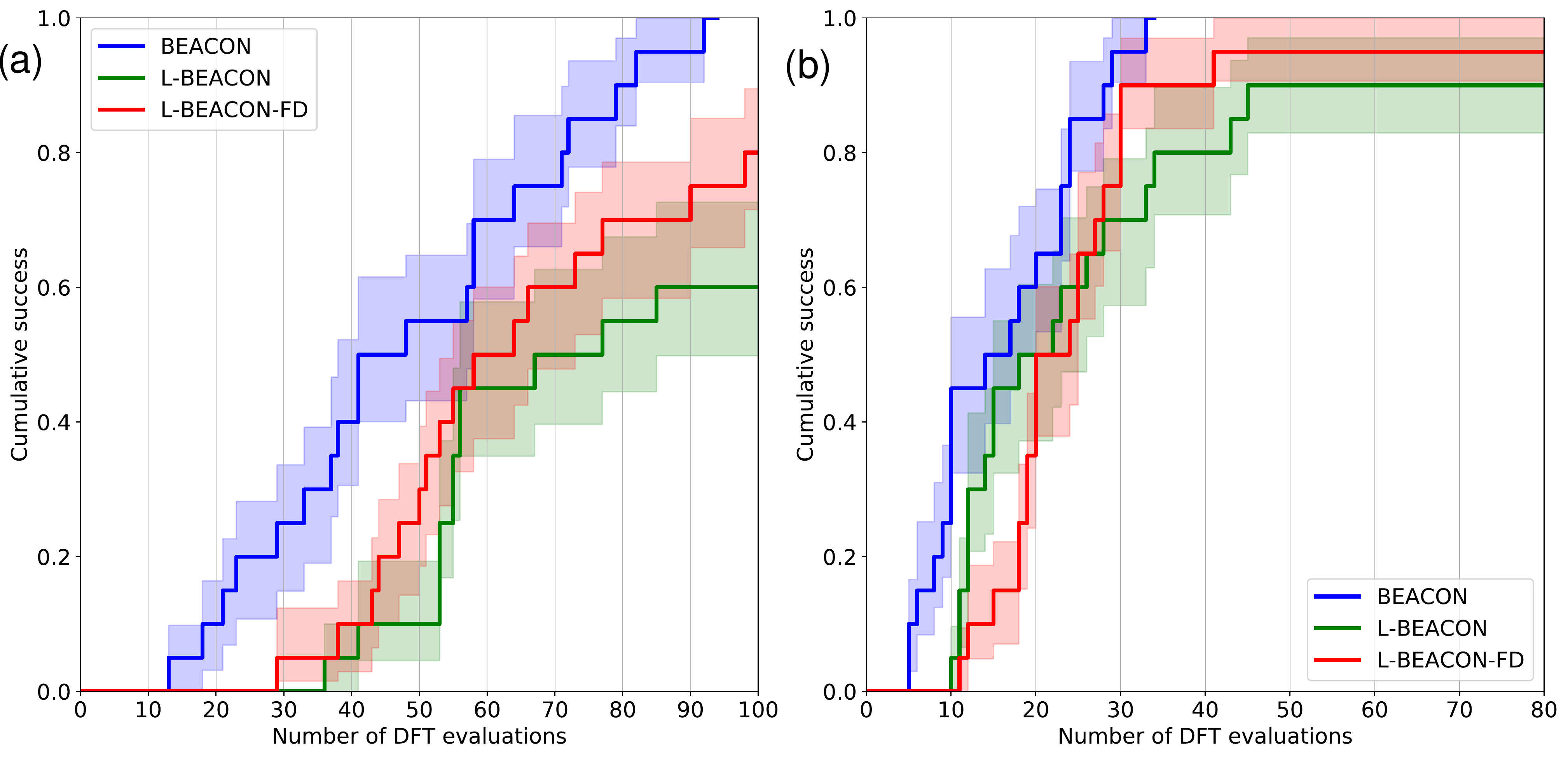}
    \caption{SiO$_2$ success curves with success thresholds of (a) 0.01 eV and (b) 0.20 eV .}
    \label{sio2_thresholds}
\end{figure}

\begin{figure}
    \centering
    \includegraphics[width=0.6\linewidth]{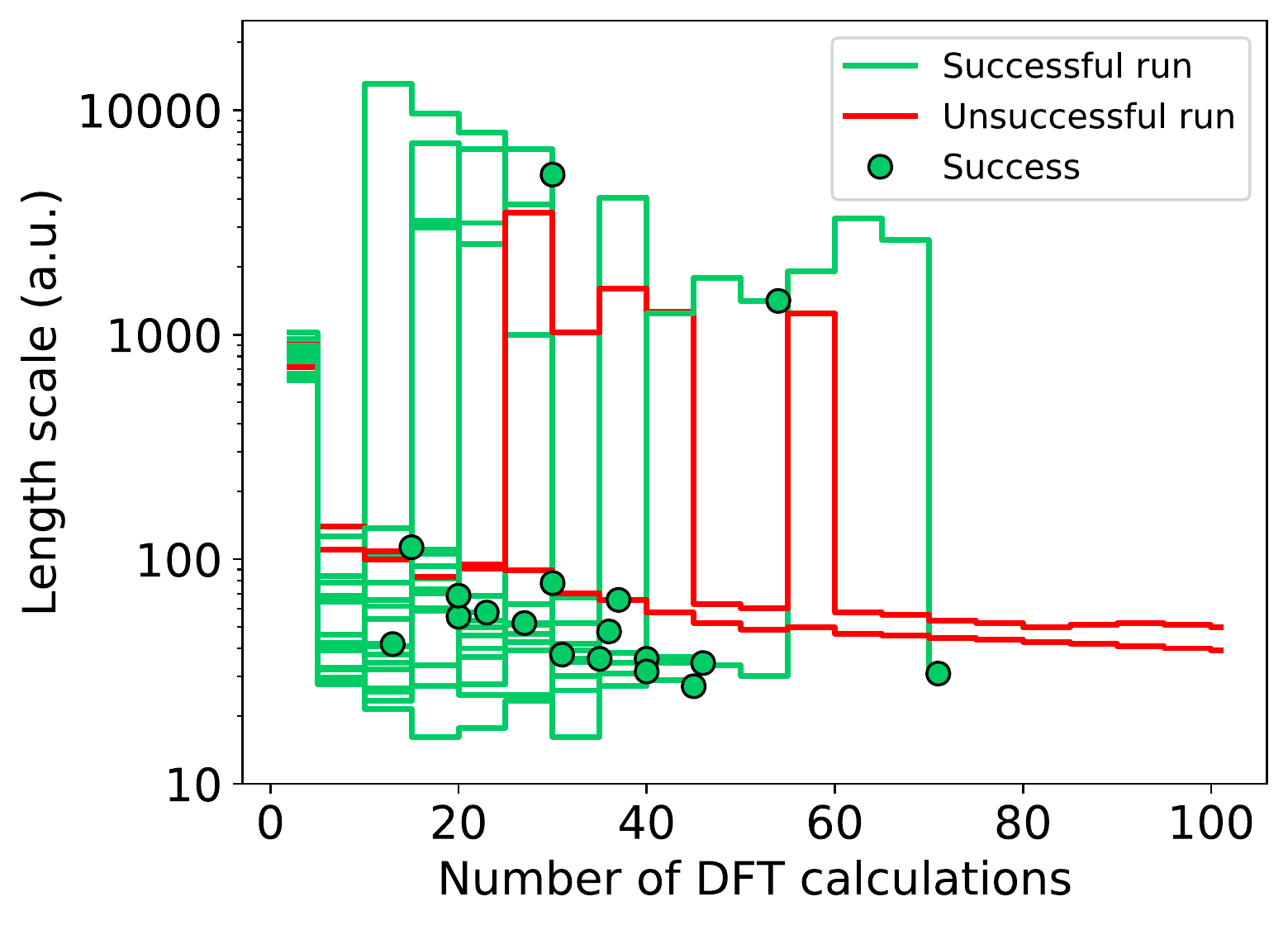}
    \caption{Fit length scales in the L-BEACON runs for SiO$_2$.}
    \label{sio2_scales_noforces}
\end{figure}

\begin{figure}
    \centering
    \includegraphics[width=\linewidth]{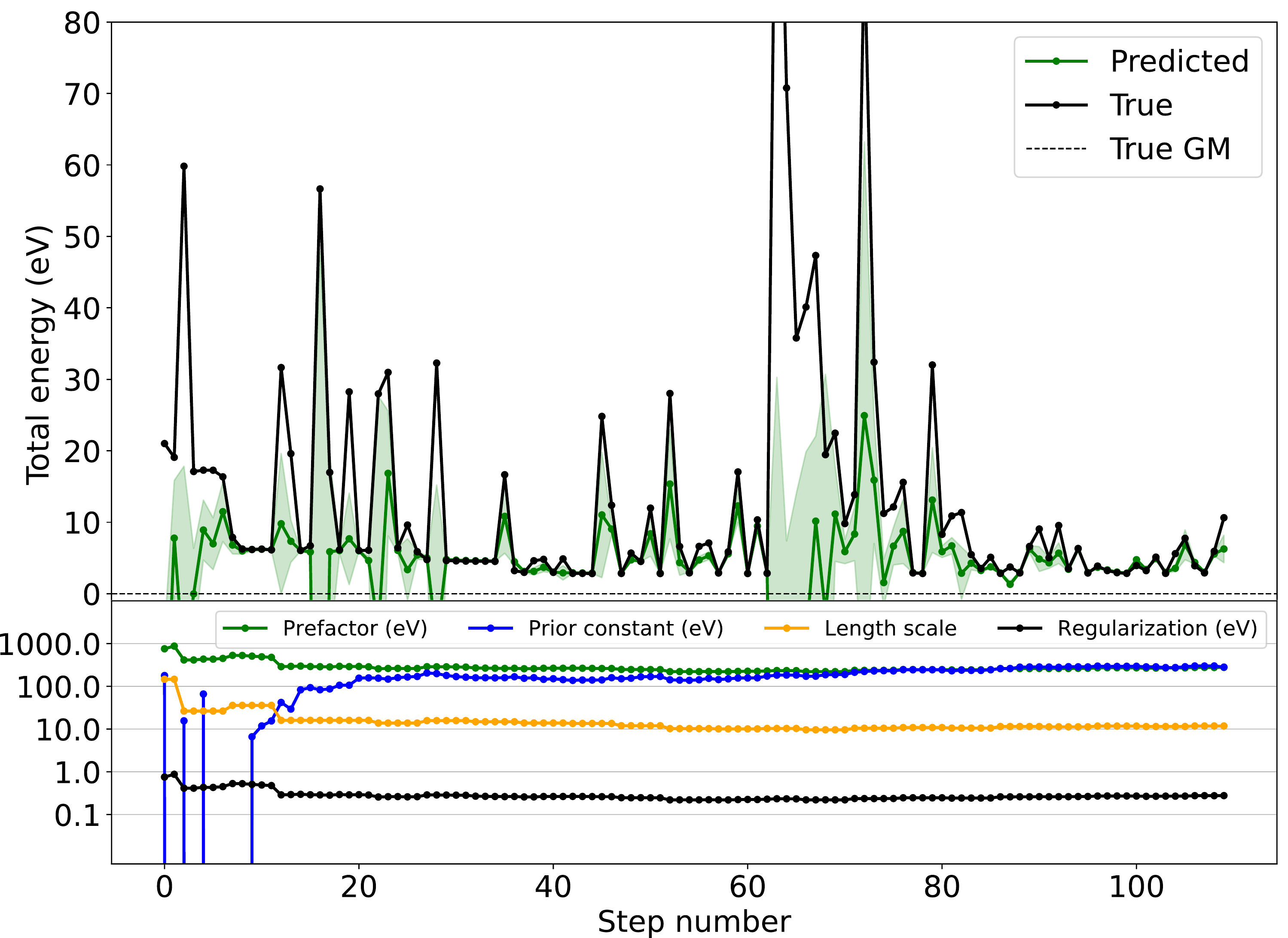}
    \caption{A single global optimization run of Ta$_6$O$_{15}$ cluster.}
    \label{ta6o15_otherrun}
\end{figure}

\pagebreak
{\bf{Computational performance of BEACON}}\\[20pt]
\begin{figure}[h]
    \centering
    \includegraphics[width=\linewidth]{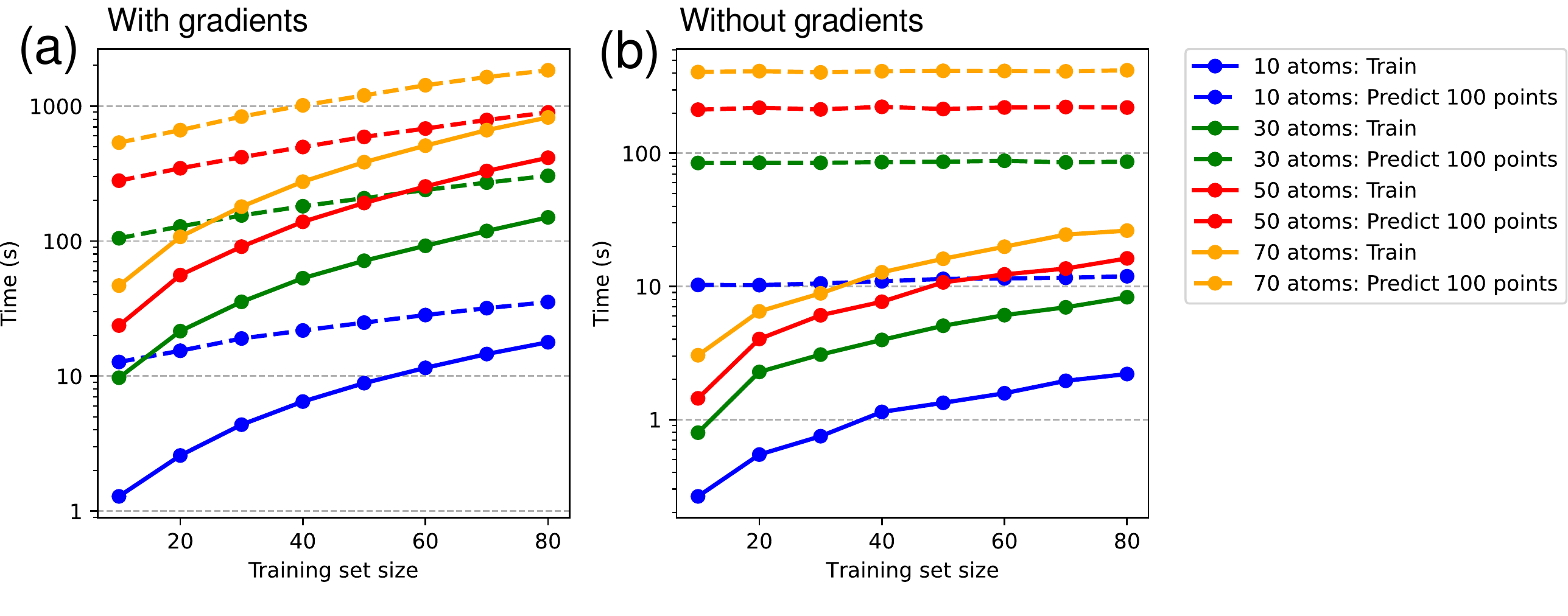}
    \caption{Computational times of training a model and predicting 100 energies of a Cu cluster with various system and training set sizes. The calculations were run on a CPU of model Intel(R) Xeon(R) CPU X5570 @ 2.93GHz / Nehalem.}
    \label{comp_times}
\end{figure}